\documentclass[12pt]{article}
\usepackage{amsmath, amsfonts, amssymb}
\usepackage{graphicx}
\usepackage{enumerate}
\usepackage{acronym}
\usepackage{natbib}
\usepackage{url} % not crucial - just used below for the URL 
\usepackage{color}
\usepackage{caption}
\usepackage{xr}
\usepackage{multirow}
\usepackage{subcaption}
\usepackage{bm}
\usepackage{theorem}
\usepackage{natbib}
\usepackage{acronym}
\usepackage{enumerate}
\usepackage{color}
\usepackage{graphicx}
\usepackage{fullpage} 

\usepackage{tikz}
\usetikzlibrary{shapes.geometric,arrows,positioning, backgrounds}
\tikzstyle{box} = [rectangle, rounded corners, minimum width=1.5cm, minimum height=0.5cm,text centered, draw=black]

\newtheorem{Lemma}{Lemma}[subsection]

\usepackage{tikz}
\usetikzlibrary{shapes.geometric,arrows,positioning, backgrounds,automata,snakes}
\tikzstyle{box} = [rectangle, rounded corners, minimum width=1.5cm, minimum height=0.5cm,text centered, draw=black]

\usepackage[all]{xy}
\usepackage{psfrag}

\usepackage{titlesec}
\titleformat*{\subsection}{\bfseries\itshape}
\titleformat*{\subsubsection}{\itshape}

\newcommand{\be}{\begin{equation}}
\newcommand{\ee}{\end{equation}}
\newcommand{\ba}{\begin{eqnarray}}
\newcommand{\ea}{\end{eqnarray}}
\newcommand{\bi}{\begin{itemize}}
\newcommand{\ei}{\end{itemize}}
\newcommand{\bn}{\begin{enumerate}}
\newcommand{\en}{\end{enumerate}}
\newcommand{\bfi}{\begin{figure}}
\newcommand{\efi}{\end{figure}}

\newcommand{\figfile}{R_Code_and_Figures/Figures}

\def \expectation  {\textrm{E}} 
\def \variance {\textrm{Var} }
\def \covariance {\textrm{Cov} }

\newcommand{\e}[1]{\ensuremath{{\rm E}[#1]}}
\newcommand{\ed}[2]{\ensuremath{{\rm E}_{#1}[#2]}}

\newcommand{\var}[1]{\ensuremath{{\rm Var}[#1]}}
\newcommand{\vard}[2]{\ensuremath{{\rm Var}_{#1}[#2]}}

\newcommand{\corr}[2]{\ensuremath{{\rm Corr}\left[#1,#2\right]}}

\newcommand{\cov}[2]{\ensuremath{{\rm Cov}[#1,#2]}}
\newcommand{\covd}[3]{\ensuremath{{\rm Cov}_{#1}\left[#2,#3\right]}}

\newcommand{\eF}[1]{\ed{\bF}{#1}}
\newcommand{\varF}[1]{\vard{\bF}{#1}}
\newcommand{\covF}[2]{\covd{\bF}{#1}{#2}}
\newcommand{\eFX}[1]{\ed{\bF \cup \bX}{#1}}
\newcommand{\varFX}[1]{\vard{\bF \cup \bX}{#1}}

\def \indicator { \mathbb{I} }
\def \identity { \boldsymbol{I} }

\def \real { \mathbb{R} }

\def \trace {\textrm {trace}}
\def \vect {\textrm {vec}}

\def \ind {\indicator}
\def \half {\frac{1}{2}}
\def \mn {\frac{m^2}{n}}

\newcommand{\mb}{\mathcal{B}}
\newcommand{\md}{\mathcal{D}}

\newcommand{\mt}{\mathcal{T}}
\newcommand{\mx}{\mathcal{X}}
\newcommand{\my}{\mathcal{Y}}

\def \bc { \mathbf{c} }

\def \bof { \mathbf{f} }
\def \bg { \mathbf{g} }

\def \bu { \mathbf{u} }
\def \bv { \boldsymbol{v} }
\def \bw { \boldsymbol{w} }
\def \bx { \mathbf{x} }
\def \by { \mathbf{y} }
\def \bz { \mathbf{z} }

\def \bB { \boldsymbol{B} }
\def \bC { \boldsymbol{C} }
\def \bF { \mathbf{F} }
\def \bG { \boldsymbol{G} }

\def \bT { \mathbf{T} }
\def \bU { \mathbf{U} }
\def \bW { \boldsymbol{W} }
\def \bX { \mathbf{X} }
\def \bY { \mathbf{Y} }

\def \bb { \bm{\beta} }
\def \bGa{ \bm{\Gamma} }
\def \bDe { \boldsymbol{\mathit{\Delta}} }

\def \beps { \bm{\epsilon} }

\def \Sig { \boldsymbol{\mathit{\Sigma}} }
\def \bth { \bm{\theta} }

\def \bTh { \boldsymbol{\mathit{\Theta}} }
\def \bO { \boldsymbol{\mathit{\Omega}} }
\def \bizero { \boldsymbol{\mathit{0}} }
\def \bzero { \mathbf{0} }
\def \mbX { \mathbb{ X } }

\def \cin { \bz } % Composite simulator input bold z.
\def \simr { \bof } % Component simulator bold f.
\def \siin { \bx } % Component simulator generic input bold x.
\def \simrin { \simr(\siin) } % Component simulator with input. 

\def \yx { \by(\bx) }
\def \yxp { \by(\bx') }
\def \yX { \by(\bX) }
\def \yXp { \by(\bX') }

\def \cX { \bc(\bX) }

\def \fX { \simr(\bX) }

\def \gX { \bg(\bX) }

\def \uX { \bu(\bX) }
\def \uXp { \bu(\bX') }
\def \vX { \bv(\bX) }

\def \wX { \bw(\bX) }

\def \disp { d }
\def \dr { \rho }
\def \ddr { h }
\def \zSM { z_{SM} }
\def \zWD { z_{WD} }
\def \zWS { z_{WS} }
\def \dispin { \disp( \cin ) }

\def \cl { \theta }

\newcommand{\seq}[2]{#1,\dots,#2}

\def \bOi { \bO^{-1}  } % Omega inverse.
\def \bDi { \bDe^{-1} }
\def \eb { \ed{\bF}{\bb} } % adjusted expectation for beta.
\def \ebt { \ed{\bF}{\bb^T} } % adjusted expectation for beta.
\def \vb { \vard{\bF}{\bb} } % adjusted variance for beta. 
\def \ewX { \e{\wX} } % expectation of w(X)
\def \vwX { \var{\wX} } % variance of w(X)
\def \WDW { \bW \, \bDe \, \bW^T }
\def \WOW { \bW^T \, \bOi \, \bW }
\def \WO { \bW^T \, \bOi }
\def \DG { \bDi \bGa }
\def \bbgls { \hat{\bb}_{GLS} }
\def \GCG { \bG^T \, \bC^{-1} \, \bG }
\def \GC { \bG^T \, \bC^{-1} }

\def \wix { \bw_i(\bX) }
\def \wjx { \bw_j(\bX) }

\newcommand{\NL}{ \nonumber \\ &&  \,\,\,\,\,\, } 
\newcommand{\NLeq}{ \nonumber \\  &=&}

\newcommand{\sam}[1]{\textcolor{black}{#1}}% SEJ comments/changes	
\usepackage[refpage]{nomencl}

\makenomenclature

\begin{document}

\title{Bayes Linear Analysis for Statistical Modelling with Uncertain Inputs}

\author{Samuel E. Jackson\footnote{samuel.e.jackson@durham.ac.uk}, David C. Woods \\
  \hspace{1cm} \\
\small Durham University, Durham, UK  \\
}

\maketitle

\normalsize

\abstract{Statistical models typically capture uncertainties in our knowledge of the corresponding real-world processes, however, it is less common for this uncertainty specification to capture uncertainty surrounding the values of the inputs to the model, which are often assumed known.
We develop general modelling methodology with uncertain inputs in the context of the Bayes linear paradigm, which involves adjustment of second-order belief specifications over all quantities of interest only, without the requirement for probabilistic specifications.  In particular, we propose an extension of commonly-employed second-order modelling assumptions to the case of uncertain inputs, with explicit implementation in the context of regression analysis, stochastic process modelling, and statistical emulation. We apply the methodology to a regression model for extracting aluminium by electrolysis, and emulation of the motivating epidemiological simulator chain to model the impact of an airborne infectious disease.}

%\vspace{-2cm}

%%%%%%%%%%%%%%%%%%%%%%%%%%%%%%%%%%%%%%%%%%%%%%

\section{Introduction \label{intro}}

% Statistical models have been successfully utilised in a wide range of scenarios over the years to make inferences about and establish connections between quantities of interest.
Most often, it is assumed that the input, or independent, variables 
of a statistical model
are known, and that uncertainties largely result from measurement error in the response and an imperfect description of the link between the independent variables and the responses.
However, in reality, the inputs may have uncertainty surrounding their values, that is, they may themselves be random variables.
This may arise as a result of imperfect measurements of the corresponding quantities in the physical experiment, or when those input variables may have been themselves modelled as an independent process.
% Modelling with uncertain inputs has been explored in the literature, and 
Such modelling is sometimes referred to as Error in Variables models \citep{durbin1954eiv, figueroa-zuniga2022rbr},
although such models are primarily concerned with estimation at specified independent variable values when the independent variables for the training data have been measured with error (often a parameterised homoscedastic error structure to be estimated).

In this paper, we develop general Bayes linear modelling methods for experiments where there is specified uncertainty in the values of the inputs for both the training data used to estimate the model and use-case data for which predictions are required. Following de Finetti \citep{TP, TP2, PVE}, Bayes linear methods \citep{BLS} involve prior second-order belief specification over all quantities of interest; the beliefs about unobserved quantities then being adjusted in light of those which have been observed.
% Such methods have been successfully applied in the literature \citep{BLSMHRH, BLAWERA}, 
A general advantage of the Bayes linear approach over the fully Bayesian approach is the lack of requirement to specify full probabilistic distributions over all quantities of interest, these often being difficult to specify meaningfully and thus often chosen for computational convenience.
Bayes linear methods have seen diverse application, including in the petrochemical \citep{BLSMHRH}, medical \citep{BLAWERA} and climate \citep{astfalck2021DRAFTcpm} sciences.

We propose an extension of second-order modelling assumptions to the case of uncertain inputs.
Whilst the methodology is generally applicable, two important 
% statistical modelling 
scenarios will be studied in detail.
Firstly, uncertain input modelling for 
% exchangeable 
regression analyses, applied to a specific example taken from \cite{goldstein1998aeb}.
In this example, input uncertainty is assumed to arise from inaccurate physical measurements training input values.
Secondly, 
% Uncertain Input Bayes Linear Emulation, which requires additional modelling assumptions in the general Bayes linear emulation setting.
we address emulation of computer models, or simulators, using a Bayes linear framework, often termed Bayes linear emulation \citep{BLUAORBMCE, Goldstein2016}.
We assume uncertain inputs arising, for example, as the outputs of a stochastic process (perhaps another emulated simulator, as is discussed in Section \ref{sec:EmulationOfSimulatorNetworks}) about which we are uncertain.
Such application first requires consideration of stochastic process modelling.
% We assume that uncertainty in the inputs is arising as a result of not knowing which values it is most appropriate to run the statistical model at, as opposed to not believing the direct correspondence between physical property values and model input values, which might from part of a further uncertainty analysis involving model discrepancy \citep{QMUCMDI, LAPP}.

The article is structured as follows.
In Section \ref{sec:BayesLinearAnalysis}, we formally introduce concepts and notation concerning Bayes linear methods, before discussing our developed general framework to statistical modelling with uncertain inputs within a Bayes linear paradigm.
 %We formally introduce concepts, notation and previous work concerning Bayes linear methods in Section~\ref{sec:BayesLinearAnalysis}.
 %In Section \ref{sec:BayesLinearModellingwithUIs}, we develop a general framework on the current Bayes linear methodology to handle statistical modelling with uncertain inputs.
 In Section \ref{sec:ExchangeableRegressions}, we demonstrate the Bayes linear modelling approach in the context of (exchangeable) regression.
 In Section \ref{sec:BayesLinearStochasticProcesses}, we develop the methodology to model stochastic processes.
 In Section \ref{UIBLE}, we combine the scenarios presented in Sections \ref{sec:ExchangeableRegressions} and \ref{sec:BayesLinearStochasticProcesses} to develop Uncertain Input Bayes Linear Emulation and demonstrate its application to emulating a chain of linked simulators.
 The methods in 
 %this 
 this section are implemented in \texttt{R}, with the developed R packages available at \url{https://github.com/Jackson-SE/UIBLE}.
 Section~\ref{conclusion} contains a brief discussion and some directions for future research.

%%%%%%%%%%%%%%%%%%%%%%%%%%%%%%%%%%%%%%%%%%%%%%

%%%%%%%%%%%%%%%%%%%%%%%%%%%%

\section{Bayes Linear Analysis and Uncertain Inputs \label{sec:BayesLinearAnalysis}}

In this article, we focus on the Bayes Linear approach \citep{LBM, BLERRM, BLS} to statistical inference, which % takes expectation as primitive, following De Finetti \citep{TP, TP2, PVE}, and 
deals with second-order belief specifications (that is, expectations, variances and covariances) of observable quantities.  Probabilities can be represented as the expectation of the corresponding indicator functions when required.

More precisely, suppose that there are two collections of random quantities, $ \mb = (B_1,...,B_r) $ and $ \md = (D_1,...,D_s) $.  Bayes linear analysis involves updating subjective beliefs about $ \mb $ given observation of $ \md $.  In order to do so, prior mean vectors and covariance matrices for $ \mb $ and $ \md $ (that is, $ \expectation[\mb] $, $ \expectation[\md] $, $ \variance[\mb] $ and $ \variance[\md] $), along with a covariance matrix between $ \mb $ and $ \md $ (that is, $ \covariance[\mb,\md] $), must be specified.  Second-order beliefs about $ \mb $ can be adjusted in the light of $ \md $ using the Bayes linear update formulae:
\begin{eqnarray} 
\expectation_\md[\mb] & = & \expectation[\mb] + \covariance[\mb,\md]\variance[\md]^{-1}(\md-\expectation[\md]), \label{UE1} \\ \variance_\md[\mb] & = & \variance[\mb] - \covariance[\mb,\md]\variance[\md]^{-1}\covariance[\md,\mb], \label{UE2} \\
\covariance_\md[\mb_1, \mb_2] & = & \covariance[\mb_1, \mb_2] - \covariance[\mb_1,\md]\variance[\md]^{-1}\covariance[\md,\mb_2]. \label{UE3}
\end{eqnarray}  
Equations \eqref{UE1}-\eqref{UE3} are the backbone of a Bayes Linear Analysis, and permit construction of Bayes linear statistical models, such as presented in the later sections of this article.
$ \expectation_\md[\mb] $ and $ \variance_\md[\mb] $ are termed the adjusted expectation and variance of $ \mb $ given $ \md $.  $ \covariance_\md[\mb_1, \mb_2] $ is termed the adjusted covariance of $ \mb_1 $ and $ \mb_2 $ given $ \md $, where $ \mb_1 $ and $ \mb_2 $ are subcollections of $ \mb $.   
% Following on from this, given a third collection of random quantities $ \ma = (A_1,...,A_t) $ we can sequentially adjust beliefs about $ \mb $ given observation of random quantities $ \md $ and $ \ma $ using a sequential Bayes linear adjustment:
% \begin{eqnarray} 
% \expectation_{\md \cup \ma}[\mb] & = & \expectation_{\md}[\mb] + \covariance_\md[\mb,\ma]\variance_\md[\ma]^{-1}(\ma-\expectation_\md[\ma]) \label{UES1} \\ \variance_{\md \cup \ma}[\mb] & = & \variance_\md[\mb] - \covariance_\md[\mb,\ma]\variance_\md[\ma]^{-1}\covariance_\md[\ma,\mb] \label{UES2} 
% \end{eqnarray}
% which adjusts the adjusted beliefs of $ \mb $ by $ \md $ now additionally by $ \ma $.
% Note that equivalent results are obtained by updating first by $ \ma $ then $ \md $ by swapping the occurrences of $ \md $ and $ \ma $ in Equations \eqref{UE1}-\eqref{UES2} above.
% Equations \eqref{UES1} and \eqref{UES2} are important for some of the discussions and calculations presented throughout Section 3 of the main text.

The Bayes linear approach differs both philosophically and practically to the fully Bayesian approach to statistical inference \citep{BLA}. 
For example, in practical problems, there are often many relevant sources of uncertainty.  A coherent fully Bayesian analysis requires specification of a full joint prior probability distribution and likelihood
to reflect beliefs about the high-dimensional structure of these uncertainties \citep{SMEPD}.  Such specification can be very difficult, hence approximations are frequently made for mathematical convenience which causes the specification to reflect some, but not all, aspects of a person's beliefs.  The theoretical coherence of the full Bayesian analysis can then get lost due to the required practical simplifications and assumptions.  Furthermore, % even if the necessary high-dimensional specifications can be adequately made, 
the resulting Bayesian analysis is often too computationally intensive to carry out in reasonable time. 
%The Bayes linear approach might be seen as either i) a practical approach or approximation to, or ii) the foundational cornerstone of, the fully Bayesian approach which removes the requirement for fully probabilistic specification of prior beliefs and data structure.
%Underlying population models are constructed by means of second-order exchangeability judgements over observable quantities, which are relatively easier to specify \citep{OMEIA}.
% The Bayes linear approach removes the requirement for fully probabilistic specification of prior beliefs and data structure.  Belief specifications are only made over observable quantities, so all belief statements can be given a direct, physical interpretation \citep{BLS}.  Underlying population models are constructed by means of second-order exchangeability judgements over observables \citep{REBSFIA, OMEIA}.  
In contrast, by only requiring belief specification up to the second order \citep{LBM}, uncertainty in model assumptions, along with any other uncertainties, can be more easily incorporated into a Bayes linear analysis.  Since linear fitting is generally computationally simpler than full conditioning, it can make for a more straightforward approach to the analysis of complex problems.
%complex problems can often be analysed in a more straightforward manner than under a full Bayesian analysis \cite{BLS}. 
% The relationship between a full Bayesian analysis and a Bayes linear analysis can be viewed in many ways.  A Bayes linear analysis may be viewed as a pragmatic approach to a full Bayesian analysis, where the task of specifying beliefs has been simplified \citep{BLA}.  Alternatively, the Bayes linear approach can be seen as the foundation of the full Bayesian approach, as is discussed in \cite{SBAPP, EBACS}.  However it is viewed, a Bayes linear analysis offers a variety of different interpretative and diagnostic tools to the full Bayesian analysis.  
% There are also similarities between the two approaches.
% For example, specifying Gaussian distributions over all quantities of interest leads to similar second-order posterior results as those given by the Bayes linear equations (\ref{UE1}) - (\ref{UE3}).  However, the interpretations of the updated quantities, and specifically the credible intervals formed, may be quite different.
For a more detailed overview and thorough treatment of Bayes linear methods, see \cite{BLA} and \cite{BLS}.
For a comparison of Bayes linear methods with the full Bayesian approach, see, for example, \cite{GFBUA}.

% OTHER CITATIONS PERHAPS IF POSSIBLE!  REDUCE CITATION e.g. 1999 GOLDSTEIN is currently refed 3 times!

% \sam{Assume $y$ is a scalar, although extension to vector easy enough?}

% \section{Bayes Linear Modelling with Uncertain Inputs \label{sec:BayesLinearModellingwithUIs}}

We now present a general statistical model 
for 
$\by(\bx) = (\seq{y_1(\bx)}{y_q(\bx)}) \in \real^q$ 
being a vector of quantities of interest assuming 
% ``controllable" (and for now, 
known
%) 
input setting $\bx \in \mbX \subseteq \real^p $.
% of the following form:
The model is given by
\be
y_i(\bx) = f_i(\bx) + \epsilon_i(\bx),  \qquad \qquad i = \seq{1}{q}. \label{stat_model}
\ee
% where $\by \in \real^q$ is a vector of real-world quantities of interest assuming ``controllable" (and for now, known) physical parameter setting $\bx \in \mbX \subseteq \real^p $, 
Here, 
$\bof(\cdot) = (\seq{f_1(\cdot)}{f_q(\cdot)})$ is a statistical model of input $\bx$, % and specific model parameters $ \bphi $,
and $ \beps(\cdot) = (\seq{\epsilon_1(\bx)}{\epsilon_q(\bx)}) $ is a residual process attempting to capture the discrepancy between $ \bof $ and $ \by$ under $\bx$.

In order to fit the model, we assume a collection of observed training data given by the $nq$-vector
$\my = \my(\mx) = ( \seq{\by_1(\mx)^T}{\by_q(\mx)^T} )^T $ corresponding to each row of the $n \times p$ matrix 
$\mx = ( \seq{\bx^{(1)}}{\bx^{(n)}} )^T $,
with 
$ \by_k(\mx), k = \seq{1}{q} $, 
being $n$-vectors of observations corresponding to the $k$th quantity $y_k$.

We aim to adjust our second-order prior belief specification about $ \by(\bx) $ across $ \mathcal{X} $ by $ \my $ using the Bayes linear update equations to obtain posterior quantities:
\be
\ed{\my}{\yx}  =    \e{\yx} + \cov{\yx}{\my} \var{\my}^{-1} (\my - \e{\my}),  \label{Exp} 
\ee
\be
\covd{\my}{\yx}{\yxp}  =   \cov{\yx}{\yxp} - \cov{\yx}{\my} \var{\my}^{-1} \cov{\my}{\yxp}\,. \label{Var}
\ee

The magnitude of the adjustment of our beliefs for $\yx$ in light of $\my$ is largely governed by the prior belief specification $\e{\yx}$ and $\cov{\yx}{\yxp}$ across $\mbX$.
For most statistical models, such prior belief specifications can be expressed as smooth functions of $\bx$ and $\bx'$, this function being determined by the model structure for $\bof$ and $\beps$ and the induced belief specification over them.
For example, 
% in Section \ref{sec:BayesLinearStochasticProcesses}, the presented covariance structure is given by 
we often consider a separable covariance structure of the form
\be
\cov{\yx}{\yxp} = c(\bx, \bx') \, \Sig,  \label{eq:sep_cov}
\ee 
where $c(\bx, \bx')$ is a stationary correlation function of $\bx$ and $\bx'$, and $\Sig$ is an output covariance matrix.  It is the structure of $\cov{\yx}{\yxp}$, and in particular correlation function $c(\bx,\bx')$, that we generalise to extend Bayes linear modelling to include uncertain inputs, firstly in general terms here, and then in specific (but broad) contexts  and examples in the subsequent sections.

% For most statistical models, $\ed{\my}{\yx}$ and $\covd{\my}{\yx}{\yxp}$ can be expressed as a function of $\bx$ and $\bx'$, this function being determined by the model structure for $\bof$ and $\beps$ and the induced belief specification over them.  It is the structure of this function that will be generalised in the next and subsequent section for Bayes linear modelling with uncertain inputs.

%%%%%%%%%%%%%%%%%%%%%%%%%%%%

Consider the generalisation of the statistical model given by Equation~\eqref{stat_model} to the following form:
\be
\by(\bX) = \bof(\bX) + \beps(\bX), \label{stat_model_UI}
\ee
where now $\bX$ is a random variable with second order belief specification $ \{ \e{\bX}, \cov{\bX}{\bX'} \} $ available for any $ \bX, \bX' $. 
Note that $y(\bX)$ is therefore the random quantity in which interest lies.

In a Bayes linear setting, we wish to obtain 
\be
\ed{\my}{\yX}  =    \e{\yX} + \cov{\yX}{\my} \var{\my}^{-1} (\my - \e{\my}),  \label{Exp_UI} 
\ee
\be
\covd{\my}{\yX}{\yXp}  =   \cov{\yX}{\yXp} - \cov{\yX}{\my} \var{\my}^{-1} \cov{\my}{\yXp}\,, \label{Var_UI}
\ee
where training data 
$ \my = \my(\mx) = ( \seq{\by_1(\mx)^T}{\by_q(\mx)^T} )^T $
are now observations corresponding to each row of 
$\mx = ( \seq{\bX^{(1)}}{\bX^{(n)}} )^T$,
each element of which is specified up to second order.
In other words, we have $ \e{\bX^{(i)}} $ and $ \cov{\bX^{(i)}}{\bX^{(j)}} $ for all $ i, j = \seq{1}{n} $.
Prior belief specification for $\my$ is given by 
\ba
\e{\my} & = & (\seq{\e{y(\bX^{(1)})}}{\e{y(\bX^{(n)})}} ), \\ 
\var{\my} & = & \{ \cov{y(\bX^{(i)})}{y(\bX^{(j)})} \}_{i,j = \seq{1}{n}} .
\ea
Appropriate methodology to obtain $ \ed{\my}{\yX} $ and $ \covd{\my}{\yX}{\yXp} $ by Equations \eqref{Exp_UI} and \eqref{Var_UI} is therefore achieved by specification of 
% by generalising the function of $\bx$ and $\bx'$ alluded to in Section \ref{sec:BayesLinearAnalysis} to be a function of second order belief specifications over $\mbX$.
% by considering 
appropriate prior belief structures $\e{\yX}$ and $\cov{\yX}{\yXp}$ 
% when $\bX$ is a random variable specified up to second order.
over any possible specification of the set \\
$\{\e{\bX}, \e{\bX'}, \cov{\bX}{\bX'}\}$.
Such belief structures often exist as generalisations of those for known $\bx, \bx'$ alluded to above.
For example, 
the separable covariance structure given by Equation \eqref{eq:sep_cov} can be generalised to
% as discussed further with specific examples in Section \ref{sec:BayesLinearStochasticProcesses}, we generalise correlation function $c(\bx,\bx')$ to
\be
\cov{\by(\bX)}{\by(\bX')} = c(\bX, \bX') \, \Sig,
\ee
with
\be
c(\bX, \bX') = c(\e{\bX}, \e{\bX'}, \cov{\bX}{\bX'}), 
\ee
implying that the correlation 
between $ \yX $ and $ \yXp $ for
two uncertain inputs $ \bX $ and $ \bX' $ is 
a function of the second order belief specification about and between the two input variables.

In the following sections, we proceed to discuss Bayes linear statistical modelling with uncertain inputs in two specific, but broad, contexts; firstly, that of regression analyses, and secondly, that of stochastic processes and emulation.

%%%%%%%%%%%%%%%%%%%%%%%%%%%%%%%%%%%%%%%%%%%%%%

%%%%%%%%%%%%%%%%%%%%%%%%%%%%%%%%%%%%%%%%%%%%%%

\section{Regression Analysis \label{sec:ExchangeableRegressions}}
 
 Consider a regression model, presented here with scalar output $ y $ for simplicity of notation and exposition:
 \be
 y(\bX) = \bX^T \bb + \epsilon(\bX), \label{eq:reg_mod}
 \ee
 where $\bb \in \real^p$ is a vector of 
 exchangeable 
 regression parameters, and $\bX \in \real^p$ is a vector random variable. 
 
 We consider a generic prior specification over $\bb$ of the form $ \e{\bb} = \bGa$ and $ \var{\bb} = \bDe $. 
 Under the usual scenario of known $ \bx $, we usually specify $ \e{\epsilon(\bx)} = 0 $ for all $\bx$, thus we view it a reasonable extension to specify $ \e{\epsilon(\bX)} = 0 $ in the random variable scenario.
 It is common for $\epsilon$ to be viewed as uncorrelated with $\bb$ \emph{a priori}, that is to specify $\cov{\bb}{\epsilon(\bx)} = 0$, so again we view it as reasonable to extend this in the random variable context to $\cov{\bb}{\epsilon(\bX)} = 0$.   
 % We must also specify a correlation structure over $\epsilon(\bX)$.  We consider two example cases of such a specification in Sections \ref{subsec:LR} and \ref{subsec:EE}, however, we first go through the calculations that don't require this distinction.
 
% Following the above prior specifications, adjusted belief specification for $y(\bX)$ can be obtained as follows:
% \ba 
% \ed{\my}{y(\bX)} & = & \e{y(\bX)} + \cov{y(\bX), \my} \, \var{\my}^{-1} \, (\my - \e{\my}) \\
% %
% \covd{\my}{y(\bX)}{y(\bX')} & = & \cov{y(\bX)}{y(\bX')} - \cov{y(\bX)}{\my} \, \var{\my}^{-1} \, \cov{\my}{y(\bX')}
% \ea
% where
% \ba
% \e{\my} & = & (\seq{\e{y(\bX^{(1)})}}{\e{y(\bX^{(n)})}} ) \\ 
% \var{\my} & = & \{ \cov{y(\bX^{(i)})}{y(\bX^{(j)})} \}_{i,j = \seq{1}{n}} .
%  \ea
% Therefore 
Following Equations \eqref{Exp_UI} and \eqref{Var_UI}, the required prior specifications are 
 $ \e{y(\bX)} $ and $ \cov{y(\bX)}{y(\bX')} $
 given any possible specification of the set
$\{\e{\bX}, \e{\bX'}, \cov{\bX}{\bX'}\}$
for random variables $\bX, \bX'$. %specified up to second-order.
 % second-order belief specifications for any $ \bX, \bX' \in \mbX $.  
 In the remainder of this section, we proceed to state results related to appropriate expression of prior beliefs of this form, both at a general level and for two specific examples of possible residual structure.  Extended derivation of these results is presented in the supplementary material.
 
We have that:
\ba
\e{y(\bX)} & = & % \e{\bX^T \bb} + \e{\epsilon(\bX)} 
% \NLeq
% %
\e{\bX^T} \, \bGa,  
\\
% \ea
% and
% \ba
\cov{y(\bX)}{y(\bX')} & = &  % \cov{\bX^T \bb + \epsilon(\bX)}{\bX'^T \bb + \epsilon(\bX')} \nonumber  \\
% %
% & = & \cov{\bX^T \bb}{\bX'^T \bb} + \cov{\epsilon(\bX)}{\epsilon(\bX')} \nonumber  \\
% %
% & & \,\, + \, \cov{\bX^T \bb}{\epsilon(\bX')} + \cov{\epsilon(\bX)}{\bX'^T \bb}
% \NLeq
% %
 \cov{\bX^T \bb}{\bX'^T \bb} + \cov{\epsilon(\bX)}{\epsilon(\bX')},  \label{eq:cov_yX}
\ea
noting that $\cov{\bX^T \bb}{\epsilon(\bX')} = 0$.
% with the final line resulting from the law of total covariance:
% \ba
% \lefteqn{\cov{\bX^T \bb}{\epsilon(\bX')}}
% \NLeq
% %
% \e{\cov{\bX^T \bb}{\epsilon(\bX') | \bX = \bx, \bX' = \bx' } } + \cov{\e{\bX^T \bb | \bX = \bx }}{\e{\epsilon(\bX') | \bX' = \bx' }}
% \NLeq
% %
% \e{\bX^T \, \cov{\bb}{\epsilon(\bX')}} + \cov{\e{\bX^T \bb | \bX = \bx }}{0}
% \NLeq
% %
% 0 + 0 = 0.
% \ea
Further, % again 
using the law of total covariance,
\ba
%\lefteqn{
\cov{\bX^T \bb}{\bX'^T \bb}
% } \\
% & = & \e{\cov{\bX^T \bb}{\bX'^T \bb | \bX = \bx, \bX' = \bx' } } + \cov{\e{\bX^T \bb | \bX = \bx }}{\e{\bX'^T \bb | \bX' = \bx' }} \nonumber \\
% & = & \e{ \bX^T \bDe \bX' } + \cov{\bX^T \e{\bb}}{\bX'^T \e{\bb}} \nonumber \\
& = & \e{ \bX^T \bDe \bX' } + \bGa^T \cov{\bX}{\bX'} \bGa.
\ea

The second term of Equation~\eqref{eq:cov_yX} is a covariance specification over the residual process $\epsilon(\bX) $.  
In Sections \ref{subsec:LR} and \ref{subsec:EE}, we present two example model belief structures in the context of statistical modelling with known inputs, demonstrating reasonable generalisation to the random variable input scenario using the ideas presented in Section \ref{sec:BayesLinearAnalysis}.

%%%%%%%%%%%%%%%%%%%%%%%%%%%%

\subsection{Linear Regression with Uncorrelated Random Error \label{subsec:LR}}

In the common known input case, the standard linear regression model with uncorrelated and homoscedastic random error can be obtained by specifying a prior covariance structure over $ \epsilon(\bx) $ of
\be
\cov{\epsilon(\bx)}{\epsilon(\bx')} = \indicator_{\bx = \bx'} \sigma^2,
\ee
for all $\bx, \bx'$,
where $ \indicator $ is the indicator function taking value $1$ if the statement is true and $0$ otherwise, and $\sigma^2$ is a common scalar variance parameter.

This specification can be simply extended to the random variable case as follows:
\be
\cov{\epsilon(\bX)}{\epsilon(\bX')} = \indicator_{\bX = \bX'} \sigma^2,
\ee
where the covariance between two random variable inputs is zero unless they are known to be the same.  Note that two random variables being the same is not equivalent to two different random variables having the same second-order belief specification.

%%%%%%%%%%%%%%%%%%%%%%%%%%%%

\subsection{Linear Regression with Correlated Error - An Example \label{subsec:EE}}

% \sam{State that $t\geq3$ in order to generalise to continuous $t$ values.}

In this section, we develop an uncertain input Bayes linear approach  for exchangeable regressions with correlated errors, building on the example of extracting aluminium by electrolysis over time presented in \cite{goldstein1998aeb}. Using the covariance structure presented in that paper for known inputs, we present a generalisation to the random variable input scenario using the ideas presented in Section \ref{sec:BayesLinearAnalysis}.

The model is of the form 
\be
y(t) = \beta_0 +  \beta_1 \, t + \epsilon(t),
\ee
and is exchangeable over $\bb$.
Note that this regression model is consistent with the general form presented in Equation~\eqref{eq:reg_mod}, with $p=2$ variables being an intercept and single controllable parameter $t$.
In \cite{goldstein1998aeb}, it is assumed that $t=\seq{1}{13}$, with a structured error model as follows:
\ba
\epsilon(t) & = & A(t) + J(t) + H(t),   \label{eps_breakdown}
\\ \nonumber
%%
%A(t) & = & \indicator_{t = t'} \sigma^2
%\\ \nonumber
%
J(t) & = & J(t-1) + Q(t),
\\ \nonumber
H(t) & = & \psi \, H(t-1) + R(t), t \geq 2,
\ea
with $ \psi \in (0,1) $, and where, for generality, we denote
$ \var{A(t)} = \sigma^2_A $,
$ \var{Q(t)} = \var{J(1)} = \sigma^2_Q $,
$ \var{R(t)} = \sigma^2_R $, 
and
$ \var{H(1)} = \sigma^2_1 $.
These terms express discrepancies from the linear trend as the sum of a pure measurement error $A(t)$, a stochastic development of the discrepancy as a random walk with drift $J(t)$, and an autoregressive term expressing the measurement of the suspended particles in the chemical analysis $H(t)$.  For further details, see \cite{goldstein1998aeb}.  Whilst the original model assumes discrete timesteps $t$, we will generalise the regression structure to continuous $t$-values, subject to $t \geq 3$ for consistency issues, before addressing the uncertain input $T$ scenario.

We assume that the three terms in Equation~\eqref{eps_breakdown} are uncorrelated, leading to
\ba
\cov{\epsilon(t)}{\epsilon(t)} & = &
% \cov{A(t) + J(t) + H(t)}{A(t') + J(t') + H(t')}
% \NLeq
% %
\cov{A(t)}{A(t')} + \cov{J(t)}{J(t')} + \cov{H(t)}{H(t')}.  \label{eq:cov_AJH}
\ea
The three terms on the right hand side of Equation \ref{eq:cov_AJH} have the following structure, where we define $d = |t-t'|$, and $t_m = \min(t,t')$, with greater exposition presented in the supplementary material:
% We consider each term of Equation~\eqref{eq:cov_AJH} separately.
% To begin with, we have
\ba
\cov{A(t)}{A(t')} & = & \indicator_{t=t'} \sigma_A^2, \label{eq:cov_A}
\\
% \ee
% Secondly, defining $d = |t-t'|$, and $t_m = \min(t,t')$, we have
% \ba
\cov{J(t)}{J(t')} & =  & % \cov{J(t_m)}{J(t_m+d)}
% \NLeq
% %
% \cov{J(t_m)}{J(t_m) + Q_{t_m+1} + ... + Q_{t_m+d}}
% \NLeq
% %
% \var{J(t_m)}
% \NLeq
% %
t_m\,\sigma^2_Q,   \label{eq:cov_J}
% \ea
% resulting in a symmetric correlation structure as follows:
% \ba
% \corr{J(t)}{J(t')} & = & \frac{\cov{J(t)}{J(t')}}{\sqrt{\var{J(t)}\var{J(t')}}}
% \NLeq
% %
% \frac{\cov{J(t_m)}{J(t_m+d)}}{\sqrt{\var{J(t_m)}\var{J(t_m+d)}}}
% \NLeq
% %
% \frac{t_m \, \sigma^2_Q}{\sqrt{t_m \,  \sigma^2_Q \, (t_m+d) \, \sigma^2_Q }}
% \NLeq
% %
% \sqrt{\frac{t_m}{t_m+d}}.  \label{eq:corr_J}
% \ea
% Similarly,
% \ba
\\
\cov{H(t)}{H(t')}  & =  & % \cov{H(t_m)}{H(t_m+d)}
% \NLeq
% %
% \cov{H(t_m)}{\psi^d\,H(t_m) + \psi^{d-1}\,R_{t_m+1} + ... + \psi\,R_{t_m+d}}
% \NLeq
% %
% \psi^d\,\var{H(t_m)}
% \NLeq
% %
\psi^d \, \left( \psi^{2(t_m-1)} \, \sigma^2_1 + \frac{1 - \psi^{2(t_m-2)}}{1 - \psi^2} \, \sigma_R^2 \right).  \label{eq:cov_H}
\ea
% since 
% \ba
% \var{H(t)} & = & \var{\psi^{t-1}\,H_1 + \psi^{t-2}\,R_2 + ... +  \psi \, R_{t-1} + R_{t}}
% \NLeq
% %
% \psi^{2(t-1)} \, \sigma^2_1 \, + \, \psi^{2(t-2)} \, \var{R_2} + ... + \psi^2 \, \var{R_{t-1}} + \var{R_{t}}
% \NLeq
% %
% \psi^{2(t-1)} \, \sigma^2_1 + ( \psi^{2(t-2)} + \psi^{2(t-3)} + ... + \psi^2 + 1 ) \, \sigma_R^2
% \NLeq
% %
% \psi^{2(t-1)} \, \sigma^2_1 + \frac{1 - \psi^{2(t-2)}}{1 - \psi^2} \, \sigma_R^2  \label{eq:var_H}
% \ea
% resulting in a symmetric correlation structure as follows:
% \ba
% \corr{H(t)}{H(t')} & = & \frac{\cov{H(t)}{H(t')}}{\sqrt{\var{H(t)}\var{H(t')}}}
% \NLeq
% %
% \frac{\cov{H(t_m)}{H(t_m+d)}}{\sqrt{\var{H(t_m)}\var{H(t_m+d)}}}
% \NLeq
% %
% \frac{\psi^d\,\var{H(t_m)}}{\sqrt{\var{H(t_m)}\var{H(t_m+d)}}}
% \NLeq
% %
% \psi^d \, \sqrt{\frac{\var{H(t_m)}}{\var{H(t_m+d)}}}
% \NLeq
% %
% \psi^d \, \sqrt{\frac{\psi^{2(t_m-1)} \, \sigma^2_1 + \frac{1 - \psi^{2(t_m-2)}}{1 - \psi^2} \, \sigma_R^2}{\psi^{2(t_m+d-1)} \, \sigma^2_1 + \frac{1 - \psi^{2(t_m+d-2)}}{1 - \psi^2} \, \sigma_R^2}}
% \NLeq
% %
% \psi^d \, \sqrt{\frac{(1 - \psi^2) \psi^{2(t_m-1)} \, \sigma^2_1 + (1 - \psi^{2(t_m-2)}) \, \sigma_R^2}{(1 - \psi^2) \psi^{2(t_m+d-1)} \, \sigma^2_1 + ( 1 - \psi^{2(t_m+d-2)})  \, \sigma_R^2}}  \label{eq:corr_H}
% \ea

Hence we have a prior residual covariance belief structure across all $t$ of
\ba
\cov{\epsilon(t)}{\epsilon(t)} & = & \indicator_{t=t'} \sigma_A^2 + t_m\,\sigma^2_Q  + \psi^d \, \left( \psi^{2(t_m-1)} \, \sigma^2_1 + \frac{1 - \psi^{2(t_m-2)}}{1 - \psi^2} \, \sigma_R^2 \right). 
\ea

We now proceed to consider suitable extensions to the uncertain input scenario.  In this case, we need an expectation and covariance structure for any possible specification of the set $\{ \e{T}, \e{T'}, \cov{T}{T'} \}$.  
We consider that the statements of expectation in the error structure simply extend to being $\e{A(T)} = 0$, $\e{J(T)} = 0$ and $\e{H(T)} = 0$; logical derivation of these statements can be shown using the law of total expectation.
Regarding covariance structure, Equation \eqref{eq:cov_A} simply extends to the following:
\be
\cov{A(T)}{A(T')} = \indicator_{T=T'} \sigma_A^2, \label{eq:covUI_A}
\ee
this being similar to the uncorrelated error term presented in Section \ref{subsec:LR}.

Equation \eqref{eq:cov_J} extends as follows, where the random variables $T_m = \min(T,T')$ and $D=|T-T'|$ correspond to known quantities $t_m$ and $d$ above:
\ba
% \lefteqn{
\cov{J(T)}{J(T')} %} 
% \NLeq
% %
% \cov{J(T_m)}{J(T_m+D)}
% \NLeq
% %
% \e{\cov{J(T_m)}{J(T_m+D) | T_m=t_m, D=d }} + \cov{\e{J(T_m)|T_m}}{\e{J(T_m+D)|T_m,D}}
% \NLeq
% %
% \e{T_m \, \sigma^2_Q} + 0
% \NLeq
%
& = & 
\sigma^2_Q \, \e{T_m}.
\ea
The tricky specification here is eliciting $ \e{T_m} $ from $ \e{T}, \e{T'} $ and $ \cov{T}{T'} $, since a function which takes the minimum of two random quantities is non-linear.  
If we know which of our random variables is bigger (without loss of generality, that $T < T'$, say), then $\e{T_m} = \e{T}$.  This is not an unrealistic situation; for example, we may have measurements at two times about which we are uncertain, but know that one of them certainly happened after the other.
% Perhaps we specify \cov{T}{T'} to be commensurate with this, or perhaps it doesn't need commenting on as it is no longer relevant.

More generally, we propose a reasonable covariance structure as follows.
% which satisfies several requirements, including;
% \bi
% \item The expected value of the covariance function assuming the conditional covariance structure as for the known input case should be an upper bound on the proposed covariance model specification.
% \item The proposed covariance structure should resolve to the covariance structure for the known input case when the inputs are indeed known (that is, when $\e{T} = t$, $\e{T'} = t'$ and $ \cov{T}{T'} = 0 $).
% \ei
Let $ \ind = \indicator_{T<T'} $ and $p = \e{\ind}$. We then condition on this event using the law of total expectation to give:
\ba
\e{T_m} & = & % \e{ \ed{\ind}{T_m}}
% \NLeq
% %
% \ed{\ind = 1}{T_m} \, \e{\ind} + \ed{\ind = 0}{T_m} \,( 1 - \e{\ind} )
% \NLeq
% %
p \, \ed{\ind = 1}{T_m} + ( 1 - p ) \, \ed{\ind = 0}{T_m}, \label{eq:Tm_Total_Exp}
\ea
where
% We have that:
\ba
\ed{\ind = 1}{T_m} & = & % \ed{\ind = 1}{T}
% \NLeq
% %
% \e{T} + \cov{T}{\ind} \, \var{\ind}^{-1} \, (\ind - \e{\ind} )
% \NLeq
% %
% \e{T} + \frac{\cov{T}{\ind}}{p(1-p)} \times (1-p) \,
% \NLeq
% %
\e{T} + \frac{\cov{T}{\ind}}{p},
\\
\ed{\ind = 0}{T_m} & = & % \ed{\ind = 0}{T'}
% \NLeq
% %
% \e{T'} + \cov{T'}{\ind} \, \var{\ind}^{-1} \, (\ind - \e{\ind} )
% \NLeq
% %
% \e{T'} + \frac{\cov{T'}{\ind}}{p(1-p)} \times (-p) \,
% \NLeq
% %
\e{T'} - \frac{\cov{T'}{\ind}}{1-p},
\ea
and 
\ba
\cov{T}{\ind} & = & % \corr{T}{\ind} \, \sqrt{\var{T} \, \var{\ind} }
% \NLeq
% %
\corr{T}{\ind} \, \sqrt{\var{T} \, p(1-p)},
\\
\cov{T'}{\ind} & = & % \corr{T'}{\ind} \, \sqrt{\var{T'} \, \var{\ind} }
% \NLeq
% %
\corr{T'}{\ind} \, \sqrt{\var{T'} \, p(1-p)},
\ea
so that:
\ba
\e{T_m} & = & % p \left( \e{T} + \frac{1}{p} \, \corr{T}{\ind} \, \sqrt{\var{T} \, p(1-p)} \right)
% \NL
% %
% + \, (1-p) \left( \e{T'} - \frac{1}{1-p} \, \corr{T'}{\ind} \, \sqrt{\var{T'} \, p(1-p)} \right)
% \NLeq
% %
p \, \e{T} + (1-p) \, \e{T'}
\NL
+ \sqrt{p(1-p)} \left( \corr{T}{\ind} \, \sqrt{\var{T}} - \corr{T'}{\ind} \, \sqrt{\var{T'}} \right).   \label{eq:eTm}
\ea

We may be able to specify values for all quantities required in Equation \eqref{eq:eTm}, in which case we have our expression for $\e{T_m}$.  If specification of $\corr{T}{\ind}$ and $\corr{T'}{\ind}$ is proving difficult, then let us first note that it is logical to assume that 
$ -1 < \corr{T}{\ind} < 0 $
(since $\ind = 1$ necessarily corresponds, on average in some sense, to smaller $T$, as in this case $T<T'$),
and
$ 0 < \corr{T'}{\ind} < 1 $, so that
\be
\e{T_m} > p \, \e{T} + (1-p) \, \e{T'} - \sqrt{p(1-p)} \left( \sqrt{\var{T}} + \sqrt{\var{T'}}  \right). \label{eq:eTm_bound}
\ee

Again, if we feel able to specify a value for $p$, then we can use it in Equation~\eqref{eq:eTm_bound}. 
Alternatively, in order to be conservative with respect to variance resolution, we can use elementary calculus to minimise the expression on the right had side of Equation~\eqref{eq:eTm_bound} over $p \in [0,1]$ to get that
\be
\e{T_m} \geq \half \left( \e{T} + \e{T'} - \sqrt{ \left( \e{T'} - \e{T} \right)^2 + \left (\sqrt{\var{T}} + \sqrt{\var{T'}} \right)^2} \right). \label{eq:Tm_lower_bound}
\ee
% and we might set
We thus propose, for $T \neq T'$, the covariance structure:
\be
\cov{J(T)}{J(T')} =   \frac{\sigma^2_Q}{2} \left( \e{T} + \e{T'} - \sqrt{ \left( \e{T'} - \e{T} \right)^2 + \left (\sqrt{\var{T}} + \sqrt{\var{T'}} \right)^2} \right).
\ee
From Equation~\eqref{eq:cov_J} with $ t = t' $, it follows that 
$ \var{J(T)} = \sigma^2_Q \, \e{T}, $
since 
$T_m = T$.
We thus make use of an indicator function in the expression for
$ \cov{J(T)}{J(T')} $
to ensure it is appropriate when it is known that $T=T'$: 
\ba
\lefteqn{\cov{J(T)}{J(T')}}
\\ & = &
\frac{\sigma^2_Q}{2} \left( \e{T} + \e{T'} - \sqrt{ \left( \e{T'} - \e{T} \right)^2 + \left( 1 - \indicator_{T=T'} \right) \left (\sqrt{\var{T}} + \sqrt{\var{T'}} \right)^2} \right) \nonumber
\ea

Note that:
\bn 
\item When 
$\var{T} = \var{T'} = 0$, 
the expression on the right hand side of Inequality~\eqref{eq:Tm_lower_bound} yields
\be
\half \left( t + t' - \sqrt{(t-t')^2} \right) = \half \left( t + t' - |t - t'| \right) = \min(t,t'), \nonumber
\ee
thus resolving to the known input scenario.
\item If it is known that $T<T'$, then, as mentioned earlier, 
% and by looking at Equation~\ref{eq:eTm} with $p = 1$, 
it would be logical to have that
$\e{T_m} = \e{T}$.
% This clearly satisfies Inequality~\eqref{eq:Tm_lower_bound}, 
Since
\be
\e{T} = \half \left( \e{T} + \e{T'} - \sqrt{ \left( \e{T'} - \e{T} \right)^2 } \right),
\ee
it is clear that $\e{T_m}$ satisfies Inequality \eqref{eq:Tm_lower_bound} in this case.
% \item We need 
% \be
% \var{J(T)} = \sigma^2_Q \, \e{T}
% \ee
% since 
% \ba
% \var{J(T)} & = & \e{\var{J(T)|T=t}} + \var{\e{J(T)|T}}
% \NLeq
% %
% \e{\sigma_Q^2 \, T} + 0
% \NLeq
% %
% \sigma^2_Q \, \e{T}
% \ea
% This doesn't directly follow from the expression on the right hand side of Inequality~\eqref{eq:Tm_lower_bound}.  
% To remedy this, we can make use of an indicator nugget term.
% For example
% \ba
% \lefteqn{\cov{J(T)}{J(T')}}
% \NLeq
% %
% \frac{\sigma^2_Q}{2} \left( \e{T} + \e{T'} - \sqrt{ \left( \e{T'} - \e{T} \right)^2 + \left( 1 - \indicator_{T=T'} \right) \left (\sqrt{\var{T}} + \sqrt{\var{T'}} \right)^2} \right) \nonumber
% \ea
\en

We extend Equation \eqref{eq:cov_H} by noticing that
\ba
\cov{H(T)}{H(T')} & \geq & % \cov{H(T_m)}{H(T_m+D)}
% \NLeq
% %
% \e{\cov{H(T_m)}{H(T_m+D)|T_m=t_m, D = d}} 
% \NL
% %
% + \cov{\e{H(T_m)|T_m=t_m}}{\e{H(T_m+D)|T_m=t_m, D = d}}
% \NLeq
% %
% \e{\psi^D\,\var{H(T_m)}} + 0
% \NLeq
% %
% \e{\psi^D \, \left( \psi^{2(T_m-1)} \, \sigma^2_1 + \frac{1 - \psi^{2(T_m-2)}}{1 - \psi^2} \, \sigma_R^2 \right)}
% \nonumber \\ & \geq &
% %
% \e{\psi^D \, \left( \psi^{2(T_m-1)} \, \sigma^2_1 + \sigma_R^2 \right)}
% \NLeq
% %
% \sigma^2_1 \, \e{\psi^{D+2T_m-2}} + \sigma^2_R \, \e{\psi^D}
% \NLeq
% %
% \sigma^2_1 \, \exp( \log( \e{\psi^{D+2T_m-2}} )) + \sigma^2_R \, \exp( \log(  \e{\psi^D} ))
% \nonumber \\ & \geq &
% %
% \sigma^2_1 \, \exp( \e{ \log( \psi^{D+2T_m-2})} ) + \sigma^2_R \, \exp(  \e{\log( \psi^D} ) )
% \NLeq
% %
% \sigma_1^2 \exp( \log \psi \, \e{D + 2T_m - 2} ) + \sigma_R^2 \exp( \log \psi \, \e{D} )
% \NLeq
% %
\sigma_1^2 \exp( \log \psi \, \e{T_M + T_m - 2} ) + \sigma_R^2 \exp( \log \psi \, \e{T_M - T_m} ), \label{eq:cov_hT}
\ea
where % the first $\geq$ is a result of the fact that, if 
% $T_m\geq3$, 
% then 
% $(1-\psi^{2(T_m-2)})/(1-\psi^2) \geq 1$,
% the second $\geq$ is a result of the fact that $\log$ is a concave function, i.e. that 
% $\e{f(X)} \leq f(\e{X})$
% if 
% $f(X) = \log(X)$.,
% and the final line makes use of the fact that 
% $ d = t_M - t_m $, where 
$ t_M = \max(t,t') $, and correspondingly
%$ D = T_M - T_m$, where
$ T_M = \max(T,T') $.

We note that $ \e{T_M+T_m} = \e{T} + \e{T'}$,
% (which we should expect and can also be derived by combining Equations~\eqref{eq:Tm_Total_Exp} and \eqref{eq:TM_Total_Exp}), 
and thus seek an upper bound for $\e{T_M - T_m}$ in order to find a lower bound for Expression~\eqref{eq:cov_hT} (since $\log\psi < 0$ if $\psi \in (0,1)$).
Similar to Equation~\eqref{eq:Tm_Total_Exp}, we use the law of total expectation on $\e{T_M}$ to get that
\ba
\e{T_M} & = & % \e{ \ed{\ind}{T_M}}
% \NLeq
% %
% \ed{\ind = 1}{T_M} \, \e{\ind} + \ed{\ind = 0}{T_M} \,( 1 - \e{\ind} )
% \NLeq
% %
p \, \ed{\ind = 1}{T_M} + ( 1 - p ) \, \ed{\ind = 0}{T_M}, \label{eq:TM_Total_Exp}
\ea
where:
\ba
\ed{\ind = 1}{T_M} & = & % \ed{\ind = 1}{T'}
% \NLeq
% %
% \e{T'} + \cov{T'}{\ind} \, \var{\ind}^{-1} \, (\ind - \e{\ind} )
% \NLeq
% %
% \e{T'} + \frac{\cov{T'}{\ind}}{p(1-p)} \times (1-p) \,
% \NLeq
% %
\e{T'} + \frac{\cov{T'}{\ind}}{p},
\\
% %
\ed{\ind = 0}{T_M} & = & % \ed{\ind = 0}{T}
% \NLeq
% %
% \e{T} + \cov{T}{\ind} \, \var{\ind}^{-1} \, (\ind - \e{\ind} )
% \NLeq
% %
% \e{T} + \frac{\cov{T}{\ind}}{p(1-p)} \times (-p) \,
% \NLeq
% %
\e{T} - \frac{\cov{T}{\ind}}{1-p}.
\ea
Combining Equations~\eqref{eq:Tm_Total_Exp} and \eqref{eq:TM_Total_Exp}, we get that
\ba
\e{T_M - T_m} & = & % \e{T_M} - \e{T_m}
% \NLeq
% %
% p\, \e{T'} + \cov{T'}{\ind} + (1-p) \, \e{T} - \cov{T}{\ind}
% \NL
% %
% - \left( p\, \e{T} + \cov{T}{\ind} + (1-p) \, \e{T'} - \cov{T'}{\ind}   \right)
% \NLeq
% %
% (1-2p) \left( \e{T} - \e{T'} \right) + 2 \left( \cov{T'}{\ind} - \cov{T}{\ind} \right)
% \NLeq
% %
% (1-2p) \left( \e{T} - \e{T'} \right) 
% \NL
% %
% + 2 \left( \corr{T'}{\ind} \sqrt{\var{T'} p(1-p)} - \corr{T}{\ind} \sqrt{\var{T} p(1-p)} \right)
% \NLeq
% %
(1-2p) \left( \e{T} - \e{T'} \right) 
\NL
+ 2 \sqrt{p(1-p)} \left( \corr{T'}{\ind} \sqrt{\var{T'}} - \corr{T}{\ind} \sqrt{\var{T}} \right).
\ea

Again, if we don't feel able to specify $\corr{T'}{\ind}$ or $\corr{T}{\ind}$, then using the assumptions that 
$ -1 < \corr{T}{\ind} < 0 $
and
$ 0 < \corr{T'}{\ind} < 1 $
we have that
\be
\e{T_M - T_m} < (1-2p) \left( \e{T} - \e{T'} \right) 
+ 2 \sqrt{p(1-p)} \left( \sqrt{\var{T'}} + \sqrt{\var{T}} \right). \label{eq:TM_Tm}
\ee

If we have specified a value for $p$, again we can use this in Expression~\eqref{eq:TM_Tm}, otherwise we can maximise over $p\in[0,1]$ to obtain 
% This follows from the calculations to bound $\e{T_m}$ above, since we note again that we have an expression of the form
% \be
% h(p) = ap + b(1-p) - c \sqrt{p(1-p)} \label{eq:hp}
% \ee
% this time with 
% $ a = - \left( \e{T} - \e{T'} \right) = \e{T'} - \e{T} $,
% $ b = \e{T} - \e{T'} $,
% and 
% $ c = -2(\sqrt{\var{T}} + \sqrt{\var{T'}}) $,
% which is maximised by Equation~\eqref{eq:hp_min_max} with $-$ instead of $+$, hence
% \ba
% h(\hat{p}) & = & \half \left( a + b - \sqrt{n} \left( \frac{c^2}{n} - \mn \right) \right)
% \NLeq
% %
% \half \left( a + b + \sqrt{n} \left( \mn - \frac{c^2}{n} \right) \right)
% \NLeq
% %
% \half \left( a + b + \sqrt{n} \, \frac{n - c^2 + c^2}{n} \right) 
% \NLeq
% %
% \half \left( a + b + \sqrt{n} \, \frac{n}{n} \right) 
% \NLeq
% %
% \half \left( a + b + \sqrt{n} \right) 
% \NLeq
% %
% \half \left( a + b + \sqrt{(b-a)^2 + c^2} \right)
% \ea
% Therefore we have that
\ba
\e{T_M - T_m} & \leq & % \half \bigg( \e{T'} - \e{T} + \e{T} - \e{T'} 
% \NL
% %
% + \, \sqrt{ \left( \e{T} - \e{T'} + \e{T} - \e{T'} \right )^2 + 4 \left( \sqrt{\var{T}} + \sqrt{\var{T'}} \right) ^2 } \bigg)  
% \NLeq
% %
\sqrt{ \left( \e{T} - \e{T'} \right )^2 + \left( \sqrt{\var{T}} + \sqrt{\var{T'}} \right) ^2 }. 
\ea

For $T \neq T' $, we therefore propose
\ba
\lefteqn{\cov{H(T)}{H(T')}}
\NLeq
\sigma_1^2 \exp( \log \psi \, ( \e{T} + \e{T'} - 2 ) ) 
\NL
+ \, \sigma_R^2 \exp \left( \log \psi \, \sqrt{ \left( \e{T} - \e{T'} \right )^2 + \left( \sqrt{\var{T}} + \sqrt{\var{T'}} \right) ^2 } \right).
\ea

% Issues as for the $J(T)$ case above are as follows:
% \bn
% \item When $\var{T} = \var{T'} = 0$, does the expression collapse to the known $\cov{H(t)}{H(t')}$ case?  No.
% \item The case when it is known that $T<T'$ doesn't seem very applicable here.
% \item What is $\var{H(T)}$?  For $J(T)$ it was very clear, and an indicator function could be used to expand the $\cov{J(T)}{J(T'}$ term to incorporate this.  Here, it isn't even clear what $\var{H(T)}$ should be.
% \en

% In my view, point (c) definitely needs addressing.  We need an expression for the variance, and it needs to coincide with the covariance expression when $T=T'$.  Point (a) may not need addressing - we could just accept that the conservative approximations/expression used above just don't resolve to the known input case, although clearly satisfy the inequalities.  Alternatively, we could make use of additional indicator functions when $\var{T} = \var{T'} = 0$ for the general covariance case and the case when $T=T'$ (i.e. recovering $\var{H(t)}$).

% So, $\var{H(T)}$ is not straightforward to access - it 
When $ T = T' $, the above expression needs to be commensurate with a reasonable expression for $ \var{H(T)} $.  Unlike $ \var{J(T)}$, we find that $ \var{H(T)} $ itself needs approximating. 
However, for a variance term we need to find an upper bound 
% (rather than a lower bound as for covariance) 
in order to be conservative with regard to prior variance specification (i.e. overestimate rather than underestimate).
We have that
\ba
\var{H(T)} & \leq & % \e{\var{H(T)|T=t}} + \var{\e{H(T)|T}}
% \NLeq
% %
% \e{\psi^{2(T-1)} \, \sigma^2_1 + \frac{1 - \psi^{2(T-2)}}{1 - \psi^2} \, \sigma_R^2}
% \NLeq
% %
% \e{\psi^{2(T-1)} \, (\sigma_1^2 - \sigma_R^2) + \frac{1 - \psi^{2(T-1)}}{1 - \psi^2} \, \sigma_R^2}
% \nonumber \\ & \leq &
\indicator_{\sigma_1^2 \leq K} K + (1 - \indicator_{\sigma_1^2 \leq K}) (\psi^4 \sigma_1^2 + \sigma_R^2),
\ea
where $K = \frac{\sigma_R^2}{1 - \psi^2}$ (see the supplementary material for justification).
% The inequality holds because as $t \rightarrow \infty$, the expression tends to $K$, but this is from above or below depending on whether the variance gets larger or smaller for larger values of $t$.  
%this inequality being justified by considering the difference between the variance at time $t$ and $t+1$:
%\ba
%\var{H(t+1)} - \var{H(t)} & = & % \psi^{2(t-1)} \sigma_R^2 - (\psi^{2(t-1)} - \psi^{2t}) \sigma_1^2
% \NLeq
% %
%  \psi^{2(t-1)} \sigma_R^2 + (\psi^{2t} - \psi^{2(t-1)}) \sigma_1^2
%  \NLeq
 %
 %\psi^{2(t-1)} (\sigma_R^2 + \psi^2 \sigma_1^2 - \sigma_1^2)
%\ea
% illustrating that $ \var{H(t)}$ is maximised at time $ t = 3 $ or as $ t \rightarrow \infty $.
% thus if $\sigma_R^2 + \psi^2 \sigma_1^2 - \sigma_1^2 \geq 0$, or in other words $\sigma_1^2 \leq K$, then the difference is positive, thus we approach the limit at infinity from below, and the expression is maximised at this limit.  On the other hand, if $\sigma_R^2 + \psi^2 \sigma_1^2 - \sigma_1^2 \leq 0$, or in other words $\sigma_1^2 \geq K$, then the difference is negative, and we approach the limit at infinity from above, in which case the expression is maximised for $t=3$ (since we enforce $t \geq 3$ at the start anyway), which yields $\psi^4 \sigma_1^2 + \sigma_R^2$.

We can now set
\ba
\lefteqn{\cov{H(T)}{H(T')}}
\NLeq
\indicator_{T=T'} L + (1 - \indicator_{T=T'}) \bigg(  \sigma_1^2 \exp( \log \psi \, ( \e{T} + \e{T'} - 2 ) ) 
\NL
+ \, \sigma_R^2 \exp \left( \log \psi \, \sqrt{ \left( \e{T} - \e{T'} \right )^2 + \left( \sqrt{\var{T}} + \sqrt{\var{T'}} \right) ^2 } \right) \bigg), \label{eq:covHT}
\ea
where 
$$ L = \indicator_{\sigma_1^2 \leq K} K + (1 - \indicator_{\sigma_1^2 \leq K}) (\psi^4 \sigma_1^2 + \sigma_R^2).$$
Note that Expression~\eqref{eq:covHT} underestimates $\cov{H(t)}{H(t')}$, as given by Equation~\eqref{eq:cov_H}, when $ \var{T} = \var{T'} = 0 $.  We do not view this as a problem, as we consider interest to lie in the case of statistical modelling with random variable input.  If it is assumed that all $t,t'$ are known, we would use Equation \eqref{eq:cov_H} throughout.

% If we wish also to use indicators to enforce the case when $\var{T} = \var{T'} = 0$ to match the case of known $t,t'$, then we can have:
% \ba
% \lefteqn{\cov{H(T)}{H(T')}}
% \NLeq
% %
% \indicator_{T=T', \var{T} \neq 0} L + (1 - \indicator_{T=T', \var{T} \neq 0}) 
% \NL
% %
% \bigg(  \sigma_1^2 \exp( \log \psi \, ( \e{T} + \e{T'} - 2 ) ) + \, \sigma_R^2 \left( \frac{1 - \psi^{2(\e{T_m}-2}}{1 - \psi^2} \right)^{\indicator_{\var{T}=\var{T'}=0}}
% \NL
% %
% \qquad \times \, \exp \left( \log \psi \, \sqrt{ \left( \e{T} - \e{T'} \right )^2 + \left( \sqrt{\var{T}} + \sqrt{\var{T'}} \right) ^2 } \right) \bigg)
% \ea

% We can show this is appropriate since 
% \bi
% \item if $T=T'$ and $\var{T} \neq 0$, then we just get $L$, the term for $\var{T}$, T being random.
% \item if $\var{T} = \var{T'} = 0$, then the expression collapses as follows:
% \ba
% \lefteqn{\cov{H(T)}{H(T')}}
% \NLeq
% %
% \sigma^2_1 \exp( \log \psi (t + t' - 2) ) + \sigma_R^2 \left( \frac{1 - \psi^{2(t_m-2)}}{1 - \psi^2} \right) \exp( \log \psi \sqrt{(t-t')^2 + 0} )
% \NLeq
% %
% \sigma^2_1 \psi^{t + t' - 2} + \sigma_R^2 \left( \frac{1 - \psi^{2(t_m-2)}}{1 - \psi^2} \right) \psi^{|t-t'|} 
% \NLeq
% %
% \psi^d \left( \sigma^2_1 \psi^{2(t_m - 1)} + \sigma_R^2 \left( \frac{1 - \psi^{2(t_m-2)}}{1 - \psi^2} \right) \right)
% \ea
% which itself naturally collapses appropriately when $t=t'$ to
% \be
% \var{H(t)} =  \sigma^2_1 \psi^{2(t - 1)} + \sigma_R^2 \left( \frac{1 - \psi^{2(t-2)}}{1 - \psi^2} \right) 
% \ee
% \item The remaining cases are as given with both indicator function values set to $0$.
% \ei

Finally, to conclude our prior specification in the case of uncertain inputs for the model presented in this section, note that, with $\bT = (1,T)^T$, we have
\ba
\e{\bT^T\bDe \bT'} & = &
% \NLeq
% %
% \trace( \bDe (\e{\bT'} \e{\bT^T} + \cov{\bT}{\bT'} ) )
% \NLeq
% %
% \trace \left( 
% \left( \begin{array}{cc} \delta_{00} & \delta_{01} \\ \delta_{10} & \delta_{11} \end{array} \right) 
% \left( \begin{array}{cc} 1 & \e{T} \\ \e{T'} & \e{T} \e{T'} \end{array} \right) 
% +
% \left( \begin{array}{cc} 0 & 0 \\ 0 & \cov{T}{T'} \end{array} \right) 
% \right)
% \NLeq
% %
% \trace \left( \begin{array}{cc} \delta_{00} + \delta_{01} \e{T'} & \delta_{00} \e{T} + \delta_{01} \e{T} \e{T'} \\ \delta_{10} + \delta_{11} \e{T'} & \delta_{10} \e{T} + \delta_{11} \e{T} \e{T'} + \cov{T}{T'} \end{array}   \right)
% \NLeq
%
\delta_{00} + \delta_{01} ( \e{T'} + \e{T} ) + \delta_{11} \e{T} \e{T'} + \cov{T}{T'},
% \ea
% and
% \ba
\\
\bGa^T \cov{\bT}{\bT'} \bGa &  = &
% \left( \begin{array}{cc} \gamma_0 & \gamma_1 \end{array} \right)
% \left( \begin{array}{cc} 0 & 0 \\ 0 & \cov{T}{T'} \end{array} \right)
% \left( \begin{array}{c} \gamma_0 \\ \gamma_1 \end{array} \right)
% = 
\gamma_1^2 \cov{T}{T'},
\ea
so that full prior specification for $y(T)$ can therefore be given by
\be
\e{y(T)} = \gamma_0 + \e{T} \gamma_1,
\ee
and 
\ba
\lefteqn{\cov{y(T)}{y(T')}}
\NLeq
\delta_{00} + \delta_{01} ( \e{T'} + \e{T} ) + \delta_{11} \e{T} \e{T'} + \cov{T}{T'} 
\NL
+ \,\, \gamma_1^2 \cov{T}{T'} 
\NL
+ \,\, \indicator_{T=T'} \sigma_A^2
\NL
+ \,\, \frac{\sigma^2_Q}{2} \left( \e{T} + \e{T'} - \sqrt{ \left( \e{T'} - \e{T} \right)^2 + \left( 1 - \indicator_{T=T'} \right) \left (\sqrt{\var{T}} + \sqrt{\var{T'}} \right)^2} \right)
\NL
+ \,\, \indicator_{T=T'} L + (1 - \indicator_{T=T'}) \bigg(  \sigma_1^2 \exp( \log \psi \, ( \e{T} + \e{T'} - 2 ) ) 
\NL
\qquad + \sigma_R^2 \exp \left( \log \psi \, \sqrt{ \left( \e{T} - \e{T'} \right )^2 + \left( \sqrt{\var{T}} + \sqrt{\var{T'}} \right) ^2 } \right) \bigg).
\ea

Given these prior specifications, we can use the Bayes linear update Equations \eqref{Exp_UI} and \eqref{Var_UI} to adjust belief specifications for any second-order specification $\{ \e{T}, \var{T} \}$ for $T$, given knowledge of system behaviour $y(\mt)$ for training set $\mt = (\seq{T^{(1)}}{T^{(n)}})^T$. Here, beliefs across $\mt$ are specified up to second order.

% \sam{phew....how much of this do we need?...I hope it was worth it!  In essence we underestimate covariance and overestimate the uncorrelated random noise term.}

%%%%%%%%%%%%%%%%%%%%%%%%%%%%%%%%%%%%%%%%%%%%%%

\section{Bayes Linear Stochastic Processes \label{sec:BayesLinearStochasticProcesses}}

We here consider modelling the relationship between input $\bx \in \real^p$ and output $\by(\bx) \in \real^q$ using a second-order stochastic process, which provides an intuitive setting for the discussion of many of our novel extensions in Uncertain Input modelling required for Bayes linear emulation in Section \ref{UIBLE}.
More precisely, we assume
\be
\by(\bx) = \bu(\bx), \label{eq:second_order_process}
\ee
with a prior assumption that 
$\e{\bu(\bx)} = 0$, 
and 
assuming that the between-outputs covariance and spatial correlation across the input space is separable, thus given by
% separable covariance structure (REFERENCES and DISCUSSION OF ASSUMPTION?) given by
\be
\cov{\bu(\bx)}{\bu(\bx')} = c(\bx, \bx') \, \Sig, \label{eq:cov_ux}
\ee
for two inputs $\bx$ and $\bx'$.
Here, $ \Sig $ is a $ q \times q $ output covariance matrix and $ c(\siin,\siin') $ is a stationary correlation function of $ \siin $ and $ \siin' $ \citep{CE, BCCM}; for example, the Gaussian correlation function for modelling a deterministic relationship (for example, that of a computer experiment) is given by
\be
c(\siin,\siin') = \exp \left \{ - \sum_{r=1}^p \left ( \frac{x_{(r)} -x'_{(r)}}{\theta_{r}} \right ) ^2 \right \}   \label{eq:GCF} ,
\ee
which depends on the specification of the correlation length parameters $ \theta_{r}, r=1,...,p $.
% \item Such a setup is a useful precursor to considering UIBLE, presented in Section \ref{UIBLE}.
% Such a setup is frequently used as a nonlinear modelling structure (REFERENCES), and 
% This example
% \sam{Partial belief information about $U$ from noisy observations, or alternative model structure, for example with nugget term for stochastic processes, could be discussed here.}

The second-order stochastic process in Equation \eqref{eq:second_order_process} can be viewed similar to a Gaussian process in the fully Bayesian paradigm with corresponding second-order specification.  Gaussian process modelling with uncertain inputs has been considered
\citep{dallaire2009lgp, mchutchon2011gpt, ye2022uqo, wang2022gpw},
however, the uncertain input location noise is often assumed to be homoscedastic and unknown, whereas we assume specification of $ \{ \e{\bX}, \e{\bX'}, \cov{\bX}{\bX'} \} $, but for the modelling structure to hold for any such specification.

In a Bayes linear setting with known inputs, the assumed covariance structure, for example as presented as separable in Equation \eqref{eq:cov_ux} and for a specific example form in Equation \eqref{eq:GCF}, makes it straightforward to adjust our  
% \item Suppose $\my = \my(\mx) = ( \seq{\by_1(\mx)^T}{\by_q(\mx)^T} )^T $, with 
% $ \by_k(\mx), k = \seq{1}{q} $ 
% being $n$-vectors of observations for the $k$th physical quantity $y_k$ corresponding to each row of the $n \times p$ matrix 
% $\mx = ( \seq{\bx^{(1)}}{\bx^{(n)}} )^T $.
second-order prior belief specification about $ \by(\bx) $ across $ \mathcal{X} $ by $ \bY $ to obtain posterior quantities, using the Bayes linear update Equations \eqref{Exp} and \eqref{Var}. 
% \be
% \ed{\bU}{\bu(\bx)}  =    \e{\bu(\bx)} + \cov{\bu(\bx)}{\bU} \var{\bU}^{-1} (\bU - \e{\bU})  
% \ee
% \be
% \vard{\bU}{\bu(\bx)}  =   \var{\bu(\bx)} - \cov{\bu(\bx)}{\bU} \var{\bU}^{-1} \cov{\bU}{\bu(\bx)}\,.
% \ee

Extension of Bayes linear stochastic process modelling to the random variable input case with second-order specification requires extension of the prior belief specifications for $\bu$, such as given by Equations \eqref{eq:cov_ux} and \eqref{eq:GCF}.  
We propose
$\e{\bu(\bX)} = 0$
and 
\be
% \cov{\bu(\bX)}{\bu(\bX')} = c(\bX, \bX') \, \Sig = c(\e{\bX}, \var{\bX}, \e{\bX'}, \var{\bX'}, \cov{\bX}{\bX'}) \, \Sig
\cov{\bu(\bX)}{\bu(\bX')} = c(\bX, \bX') \, \Sig \label{eq:covuXuX}
\ee
where
\be
c(\bX, \bX') = c(\e{\bX}, \e{\bX'}, \cov{\bX}{\bX'}) \label{eq:cXX} % \, \Sig
\ee
which implies that the correlation 
between $ \uX $ and $ \uXp $ for
two uncertain inputs $ \bX $ and $ \bX' $ is 
a function of the second order belief specification about and between the two input variables.
As an example, we propose the following extension to the Gaussian correlation function, given by Equation \eqref{eq:GCF}:
\be\label{UIcor}
\begin{split}
c(\bX,\bX') & = \exp \left\{ - \e{(\bX - \bX')^T\bTh^{-2}(\bX - \bX')}\right\} \\
 & = \exp \left\{ - \sum_{r=1}^p \left( \frac{\expectation[X_{(r)} - X'_{(r)}]^2 + \variance[X_{(r)} - X'_{(r)}]}{\theta^2_r} \right) \right\} \,,
\end{split}
\ee
with positive-definite diagonal matrix $\bTh^{-2}$ having entries $(\Theta^{-2})_{rr} = 1/\theta_r^2$, and the second line obtained from standard results on the expected value of a quadratic form \citep[][pp. 200-201]{LMRDMA}.  
This choice satisfies the desirable property that it reduces to a standard form of correlation function if $ (\bX, \bX') = (\siin, \siin') $ are known.
We also derive two further important results for this new correlation function in the form of two lemmas, proofs of which can be found in the supplementary material.
\begin{Lemma}\label{pdlemma}
For random variables $\bX, \bX'$ with finite first and second moments, the kernel function $c(\bX, \bX')$ from~\eqref{UIcor} is positive-definite.
\end{Lemma}

\begin{Lemma}\label{lblemma}
Covariance function~\eqref{eq:cXX}, with $c(\bX, \bX')$ given by~\eqref{UIcor}, is a lower bound  under the Loewner (partial) ordering on the covariance obtained by assuming the conditional covariance
\be \label{condcov}
\cov{\uX}{\uXp\,|\,\bX = \bx, \bX'=\bx'} = \exp \left\{ - \sum_{r=1}^p \left( \frac{x_{(r)} - x'_{(r)}}{\theta_r} \right)^2  \right\} \, \Sig\,.
\ee
Moreover, the $kk$th element of~\eqref{UIcor} is a lower bound on the expected value, with respect to $\bX, \bX'$, of the $kk$th element of~\eqref{condcov} ($k=1,\ldots q$).
\end{Lemma}
Whilst the proposed correlation function of Equation \eqref{UIcor} can be viewed simply as a modelling assumption, Lemma \ref{lblemma} shows that it can also be derived as an approximation to the conditional covariance between $ \bu(\bX) $ and $ \bu(\bX') $ assuming the standard Gaussian correlation function that one might use for known $ \bx, \bx' $.  
In terms of 
% an emulator
a Bayes linear adjustment, this quantity reflects the amount of resolved uncertainty given the training 
% runs, 
data, hence an underestimation (lower bound) of this quantity is preferable to an overestimation.  
Similar derivations for other correlation function forms commonly presented in the literature (for example, given by \citealp{DPGP}), is an avenue for future research.
% \sam{We could add in a 1D example here, with zero-mean prior specification for example.}

%%%%%%%%%%%%%%%%%%%%%%%%%%%%

%%%%%%%%%%%%%%%%%%%%%%%%%%%%

%%%%%%%%%%%%%%%%%%%%%%%%%%%%%%%%%%%%%%%%%%%%%%

\section{Uncertain Input Bayes Linear Emulation \label{UIBLE}}

We extend uncertain input modelling methodology to Bayes linear emulation, which can in some sense be viewed as combining the applications of Sections \ref{sec:ExchangeableRegressions} and \ref{sec:BayesLinearStochasticProcesses} together.
Emulators are typically utilised as statistical approximations of mathematical models, represented as computer code, or simulators, of scientific processes.
Such mathematical models encapsulate the key features of the system and facilitate prediction and decision making.
Emulation is often required as a result of the computational expense of a typical simulator leading to substantial output uncertainty across the input space due to the small number of input combinations for which it is feasible to run the simulator.
The difficulties encountered as a result of the overall modelling uncertainties are exacerbated in the case of Uncertain Inputs, as discussed in this section. 
We begin with a review of Bayes linear emulation.

%%%%%%%%%%%%%%%%%%%%%%%%%%%%

\subsection{Bayes Linear Emulation\label{sec:BLE}}

We assume system behaviour of interest $ \by $ is modelled by a generic deterministic simulator $\bof$.
The simulator has input vector $ \siin = ( \seq{x_{(1)}}{x_{(p)}} ) \in \mathbb{X} \subseteq \real^p $,  and outputs vector 
$ \simrin = ( \seq{f_1(\bx)}{f_q(\bx)} ) \in \simr(\mathbb{X}) \subseteq \real^q $. 
% We define a Bayes linear emulator to be a meta-model which expresses beliefs 
We represent our beliefs about the behaviour of 
% any scalar simulator output component 
$ \simrin $
% , for any input $ x $, 
in the following form \citep{PFTIMMPS}: 
\be
\simrin = \bg(\siin)^T \bB + \bu(\siin) % = \sum_{v=1}^{m}  t_{v}(\siin) \bb_{.v} + \bu(\siin) 
\, ,  \label{emulator}
\ee
where $ \bg(\siin) $ is an $ m $-vector of known basis regression functions, $ \bB $ is an $ m \times q $ matrix of unknown regression coefficients, and $ \bu(\siin) $ is a $ q $-dimensional second-order weakly-stationary stochastic process.  

Let $\bb = \vect{(\bB)}$ be an $mq$-vector resulting from stacking the columns of $\bB$, with a generic prior specification $ \e{\bb} = \bGa $ and $ \var{\bb} = \bDe $.
We also make the common assumptions that $ \e{\bu(\siin)} = \bzero $, $ \cov{\bb}{\bu(\siin)} = \bizero $, 
% with $\bb = \vect{(\bB)}$ a $mq$-vector resulting from stacking the columns of $\bB$, 
and covariance between $ \bu(\siin) $ and $ \bu(\siin') $ is of the separable form given by Equation \eqref{eq:cov_ux}.

Consistent with previous notation, suppose $ \bF = \simr(\mx) = ( \seq{\simr_1(\mx)^T}{\simr_q(\mx)^T} )^T $ is an $ nq $-vector with  $ \simr_k(\mx) $ being $n$-vectors of simulator output $ k = \seq{1}{q} $ run at each row of the $ n \times p $ design matrix $ \mx = ( \seq{\bx^{(1)}}{\bx^{(n)}} )^T $.
We can adjust our second-order prior belief specification about $ \simr(\bx) $ across $ \mathcal{X} $ by $ \bF $ using the Bayes linear update Equations \eqref{BLExp} and \eqref{BLVar}, now with $\by$ replaced with $\bof$, to obtain posterior quantities:
\be
\eF{\simrin}  =    \e{\simrin} + \cov{\simrin}{\bF} \var{\bF}^{-1} (\bF - \e{\bF}),  \label{BLExp} 
\ee
\be
\varF{\simrin}  =   \var{\simrin} - \cov{\simrin}{\bF} \var{\bF}^{-1} \cov{\bF}{\simrin}\,. \label{BLVar}
\ee

\subsection{Bayes Linear Emulation with Uncertain Inputs}

Uncertain Input Bayes Linear Emulation (UIBLE) is a computationally efficient extension to Bayes linear emulation in the case of uncertain inputs.
We consider an emulator setup for $ \simr $ similar to that discussed in Section \ref{sec:BLE}. 
Following Equation \eqref{emulator}, we choose to decompose the vector of training runs $ \bF $ as follows:
\be
\bF = \vect(\bG \bB) + \bU = \bW \bb  + \bU, \nonumber
\ee
where 
$ \bG = ( \seq{ \bg(\siin^{(1)}) }{ \bg(\siin^{(n)})})^T $ is an $ n \times m $ matrix of regressors at the known design points in $ \mx $,  $ \bW = \identity_q \otimes \bG $, $ \identity_q $ is a $ q \times q $ identity matrix, $ \otimes $ represents the kronecker product, $ \bU = \bu(\mx) $ is an $ nq $ vector of residuals, and recall $ \bb = \vect( \bB ) $
with prior specification $ \e{\bb} = \bGa $ and $ \var{\bb} = \bDe $.

We wish to make inference about $ \fX $, where $ \bX \in \real^p $ is an uncertain (random variable) input to $ \simr $.  
Following Equation \eqref{emulator}, $ \fX $ can be written as 
\be
\fX = \gX^T \bB + \uX = \wX \bb + \uX, \nonumber
\ee
where $ \wX = \identity_q \otimes \gX^T $.
We 
assume $ \e{\uX} = \bzero $ and $ \cov{\bb}{\uX} = \mathbf{0} $, these assumptions being a reasonable extension to the prior specification that may be made in the case of known inputs \citep{DPSEHMM}.
We also utilise the covariance structure for $\bu(\bX)$ as given by Equations \eqref{eq:covuXuX} and \eqref{eq:cXX}.
We therefore have that $ \e{\bU} = \bzero $, $ \var{\bU} = \bO = \Sig \otimes \bC  $ and $ \cov{\bb}{\bU} = \bizero $, where we define
\be
\bC = \left( \begin{array}{cccc} c(\siin_1, \siin_1) & c(\siin_1, \siin_2) & \cdots & c(\siin_1, \siin_n) \\ c(\siin_2, \siin_1) & c(\siin_2, \siin_2) & \cdots & c(\siin_2, \siin_n) \\ \vdots & \vdots & \ddots & \vdots \\ c(\siin_n, \siin_1)& c(\siin_n, \siin_2) & \cdots & c(\siin_n, \siin_n) \end{array} \right)  \nonumber .
\ee
We also define $ \vX = \Sig \otimes \cX $ and $ \cX = ( \seq{ c(\bX, \siin_1) }{ c(\bX, \siin_n) } ) $.
We now proceed to state the adjusted belief formulae for $ \fX $ by $ \bF $ in the form of two lemmas, proofs of which can be found in the supplementary material. 

\begin{Lemma}\label{lem1} 
The expected value of $ \simr(\bX) $, adjusted by $ \bF $, is given by:
\be
\eF{\fX} = \ewX \, \eb   +    \vX \, \bOi \, (\bF - \bW \, \eb ) \,. \label{UIBLE_lem_E}
\ee
\end{Lemma}

\begin{Lemma} \label{lem2} 
The variance of $ \simr(\bX) $, adjusted by $ \bF $, is given by:
\ba
\varF{ \fX } & = & \e{ \wX \, \vb \, \wX^T }   +   \ebt \, \vwX \, \eb + \, \Sig  
\NL
\,    - \,   \vX \, \bOi \, \vX^T    +    \vX \, \bOi \, \bW \, \vb  \, \bW^T \, \bOi \, \vX^T
\NL
\, - \, \ewX \, \vb \, \bW \, \bOi \, \vX^T    
\NL
\, - \,   ( \ewX \, \vb \, \bW \, \bOi \, \vX^T ) ^T \,. \label{UIBLE_lem_Var}
\ea
\end{Lemma}  

Specification of $  \e{\wX} $ and $ \var{\wX} $ is straight forward for first-order linear regression functions.  
It is also possible for further functions of the input components, but these transformed input components will require a sensible second-order specification.
Expressions for $\eb$ and $\vb$ (extended derivations for which are provided in the supplementary material) are given by
\ba
\eb & = & ( \WOW + \bDi )^{-1} ( \WOW \, \bbgls   +   \DG ), \\
\vb & = & (\WOW + \bDi)^{-1},
\ea
where
\be
\bbgls  =  \identity_q \otimes ( \GCG )^{-1} \GC \, \bF.
\ee
As for the common known input case, vague priors on $ \bb $, which we define to mean that the eigenvectors of $\bDe$ tend to $\infty$ and thus the prior has negligible effect on the posterior, resulting in $ \eb = \bbgls $ and $ \vb = (\WOW)^{-1} $.
% $ \vb = \identity_q \otimes ( \bG^T \, \bC^{-1} \, \bG )^{-1}   $.  

Assuming an appropriate correlation structure across any possible specification of the set $\{ \e{\bX}, \e{\bX'}, \cov{\bX}{\bX'} \}$, the results of Lemmas \ref{lem1} and \ref{lem2} can be used to provide a second-order approximation of the output ($\eF{\fX}$ and $\varF{ \fX }$) of any simulator at random variable input $\bX$ for which a second-order belief specification $\{ \e{\bX}, \var{\bX} \}$ is itself provided.

%%%%%%%%%%%%%%%%%%%%%%%%%%%%

\subsection{Emulation of Simulator Networks \label{sec:EmulationOfSimulatorNetworks}}

An application of UIBLE is that of the Bayes linear emulation of simulator networks.

Complex systems can often be most appropriately modelled as a network of simpler \textit{component} simulators, that together form a \textit{composite} simulator of the entire system of interest. 
One simple, but important, example from epidemiology, that will be considered further in Section \ref{sec:DDR}, combines an atmospheric Anthrax dispersion simulator \citep{legrand2009etl}, labelled $ \disp(\cdot) $, with a dose-response (DR) simulator \citep{groer1978drc}, labelled $ \dr(\cdot) $, in a simple \textit{chain} network; see Figure~\ref{GMDDR}. The composite dispersion dose-response (DDR) simulator $ \ddr = \dr( \disp( \cin ) ) $ models the overall process, where $ \cin $ can be viewed as the input to $ \disp $ or $ \ddr $.

Utilising simulator networks for uncertainty quantification is challenging, largely due to the variety of sources of uncertainty for each individual component simulator \citep{AMA} and the necessity of propagating that uncertainty through the network. 
The issues surrounding computational expense, discussed in the introductory paragraph to Section \ref{UIBLE}, are also exacerbated, requiring use of emulation.
A key question in such scenarios is whether combining emulators for the component simulators within a network can result in more powerful approximations than emulating the network as a single composite simulator.

The output of a Bayes linear emulator, or UIBLE, as given by Equations \eqref{BLExp} and \eqref{BLVar}, or \eqref{UIBLE_lem_E} and \eqref{UIBLE_lem_Var} respectively, is a second order belief specification across the output components of $\bof(\bx)$, which can subsequently be used as the second order belief specification of the input to a subsequent simulator or emulator.
As a result, UIBLE can be used to approximate arbitrarily large networks of simulators, where the random input $ \bX $ to one simulator is taken to have a second-order belief specification arising from a previous emulator.

% It should be noted that 
Emulation of simulator networks has been addressed elsewhere in the literature.  
\cite{CCMLSE} proposed coupling two simulators by linking independently developed Gaussian process emulators of the simulators. 
Their motivation arose from potentially having \textit{separate} training runs for two simulators $\simr^1 = $ \emph{bent} and $\simr^2 = $ \emph{puff}, where \emph{bent} simulates volcanic ash plumes arising from a vent and \emph{puff} simulates ash dispersion. As a result, direct emulation (similar to that discussed in Section~\ref{sec:BLE}) of the composite simulator defined by the chain is not possible.  For specific Gaussian process forms, they derived closed form expressions for the overall mean and variance 
arising from linking the two emulators and applied these quantities within a normal distribution approximation for the composite emulator.
Subsequently, \cite{ming2021lgp} extended the work of emulating coupled simulators to much larger networks, also deriving closed-form mean and variance expressions when using a class of Mat\'{e}rn correlation functions.
The availability of second-order posterior statistics for the emulation of simulator networks provided in this literature, but the lack of a closed-form distribution, naturally suggests the use of Bayes linear approaches in the same context, hence the application of UIBLE.

Emulation, particularly Gaussian process emulation, of simulator networks, can also be compared to emulation using deep Gaussian processes \citep{DGP, HDADGP}.  Deep Gaussian processes arise from belief networks about simulator behaviour based on Gaussian process mappings, such that layers of Gaussian process latent variables exist between simulator input and output, these being marginalised variationally (see, for example, \citealp{titsias2010bgp}).
Whilst similar, the intermediate variables in a simulator network represent physical system properties, which aids the construction and modelling
of the emulators for each of the component processes.  
Direct use of a deep Gaussian process for the entire network will not exploit this additional information.
Linking deep Gaussian processes of component simulators can make use of the advantages of both observable and latent variables \citep{ming2022DRAFTdgp}, however, extension of such strategies to the Bayes linear paradigm, where one is usually concerned with belief specification over observable quantities, is an area for future research.

%%%%%%%%%%%%%%%%%%%%%%%%%%%%

%%%%%%%%%%%%%%%%%%%%%%%%%%%%

\subsection{Application to a DDR Chain of Simulators \label{sec:DDR}}

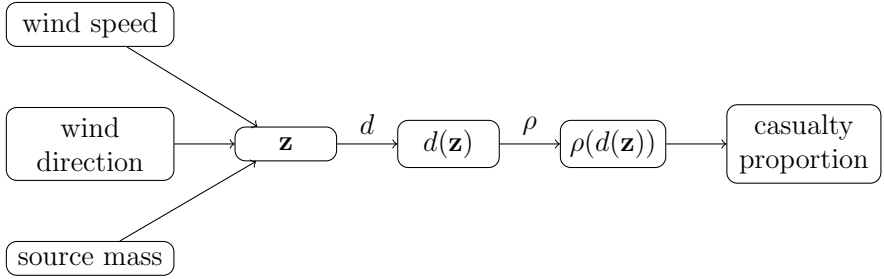
\begin{figure} 
 \begin{center}
\begin{tikzpicture}[node distance=0.8cm, scale=0.9, every node/.style={scale=0.9}]

    \node [box, text width =2.2cm]              (a1) {wind speed};
    \node [box, below=of a1, text width = 2.2cm] (a2) {wind direction};
    \node [box, below=of a2, text width = 2.2cm] (a3) {source mass};
    \node [box, right=of a2] (a4) {$\cin$};
    \node [box, right=of a4] (a5) {$\disp(\cin)$};
     \node [box, right=of a5] (a8) {$\dr(\disp(\cin))$};
    \node [box, right=of a8, text width = 2cm] (a9) {casualty proportion};
     \draw [->] (a1) -- (a4);
     \draw [->] (a2) -- (a4);
     \draw [->] (a3) -- (a4);
    \draw [->] (a4) -- (a5) node[midway, above] {$\disp$};
     \draw [->] (a5) -- (a8) node[midway, above] {$\dr$};
     \draw [->] (a8) -- (a9);

\end{tikzpicture}
  \caption{Graphical representation of the DDR 
 network of simulators $ \ddr( \cin ) = \dr( \disp( \cin ) ) $%, where $ \disp $ represents the dispersion model, and $ \dr $ represents the DR model
 . 
\label{GMDDR}}
\end{center}
\end{figure}

In this section, we apply UIBLE to an important example from epidemiology which combines an atmospheric Anthrax dispersion simulator \citep{legrand2009etl}, labelled $ \disp(\cdot) $, with a dose-response (DR) simulator \citep{groer1978drc}, labelled $ \dr(\cdot) $, in a simple \textit{chain} network; see Figure~\ref{GMDDR}.

The dispersion simulator models the spread of a released biological agent across a given spatial domain, with input parameters of interest corresponding to physical quantities wind speed ($ \zWS $), wind direction ($ \zWD $) and source mass ($ \zSM $). Simulator outputs $ \dispin $ represent dose at each location across the domain, however, we consider a single spatial location output for illustrative purposes.
Due to the behaviour of $ \disp(\cdot) $,  we chose to emulate a transformation of the output, namely $ f^1(\siin_1) = \log(\disp(\siin_1) + 1) $, treating this transformed function $ f^1(\cdot) $ as the first simulator of the network.

The DR simulator takes dose, $x$, as input, and outputs casualties, $ \rho(x)$, as a proportion of the population at a particular spatial location.
However, to be consistent with $ f^1(\cdot) $, we consider the second simulator 
to be $ f^2(x_2)  = \dr( \exp( x_2 ) - 1 ) $, so that $ h(\cin) = f^2(f^1(\cin)) = \dr( \disp( \cin ) ) $.
We also note that whilst $ \disp(\cdot) $ is computationally expensive, dose-response model $ \rho(\cdot) $ is not; however we emulate both simulators to demonstrate the efficacy of our methods. Our methods are also applicable and effective when only a subset of the simulators in a network require emulation. As $ f^2(\cdot) $ here is straightforward to emulate, our application also effectively demonstrates  the use of the methods for this special case. 
The simplicity of this second simulator additionally highlights a key benefit of the emulation of simulator networks methodology in general, since if the composite simulator is emulated as a single model, neither the computational efficiency of running the second simulation nor the simplicity of the behaviour of this simulator can be exploited.  

The composite simulator $ h = f^2 \cdot f^1 = \rho \cdot d $ takes wind speed, wind direction and source mass as input $ \cin $, and directly outputs a proportion of casualties $ h(\bz) $.  
% The DAG of this setup can be presented as that on the right of Figure \ref{DAGs}, with $ \simr^1 $ and $ \simr^2 $ (now scalar output) as discussed above.
An expanded DAG showing 
Figure \ref{GMDDR} shows the links between the original simulators $ \disp $ and $ \dr $, their inputs, output and corresponding physical quantities.

The primary interest of decision makers is the impact of release conditions on the proportion of casualties. This assessment requires linking the two component simulators, each of which implements modelling from two different groups of experts.
To address the question of whether emulating individual simulators in the network is preferable to a single emulator of the composite simulator, we will compare direct Bayes linear emulation of $h(\cdot)$ with application of UIBLE to $f^2(\cdot)$ given standard Bayes linear emulation of $f^1(\cdot)$.

We proceeded to construct Bayes linear emulators for each of the component simulators $ f^1 $ and $ f^2 $, as well as the composite simulator $ h $. Ranges of interest of the inputs to simulator $ f^1 $ (and thus $ h $) are 
% as given in Table \ref{IR_f}, 
$ (z_{WD}, z_{WS}, z_{SM} ) \in [37,63]^\circ \times [1,150]\textrm{ms}^{-1} \times [0.001,1]\textrm{kg} $,
each of which were scaled to $ [-1,1] $ for the purposes of our analysis.  We constructed a training point design for $ f^1 $ and $ h $ using a maximin Latin hypercube \citep{CTM, BPDF} of size 50 across the three input dimensions.  In contrast, simulator $ f^2 $ is one-dimensional, thus the need for fewer training points, so we take a random sample of 20 points from a uniform distribution.  This idea of using independent space-filling designs for the construction of each of the two emulators is similar to that proposed in \cite{CCMLSE}.  More sophisticated design strategies are discussed in Section \ref{conclusion}, however, yield much scope for future research.
In order to explore the effects of design size, we subsequently repeated the analysis using 30 training points for each simulator, this following the ad-hoc $10d$ rule-of-thumb \citep{CSSCE}.
It is correct to use 30 (as opposed to 10) training points even for $f^2$, since it is the power of the component-wise emulation of simulator networks that permits separation of the 1-dimensional emulator from the 3-dimensional composite simulator.
% Note that, as a result of needing to run simulator $ f^2 $ many times at each training point in order to find the mean function, this simulator is slow, hence if fewer training points are required to train the emulator for this simulator it is already an advantage.  
For each of the emulators for $ f^1,f^2 $ and $ h $, we assumed a Gaussian correlation function, as given by Equation \eqref{eq:GCF}, along with a first-order polynomial mean function.  We represent the scalar variance parameter and correlation length vectors as $ \sigma^2_1, \sigma^2_2, \sigma^2_h $ and $ \bth_1, \cl_2, \bth_h $ respectively. We fit these parameters using maximum likelihood for each emulator, this permitting a fair comparison between the emulation methods presented.  
It should be noted that whilst it can be argued that use of maximum likelihood lies outside of the Bayes linear paradigm in which we present the methodology of this article, it provides a useful tool for obtaining valid emulators.  Alternative approaches, such as making use of cross-validation, might also be used.

\begin{figure}
\centering
\includegraphics[height=12.5cm, width = 13cm, angle = 0]{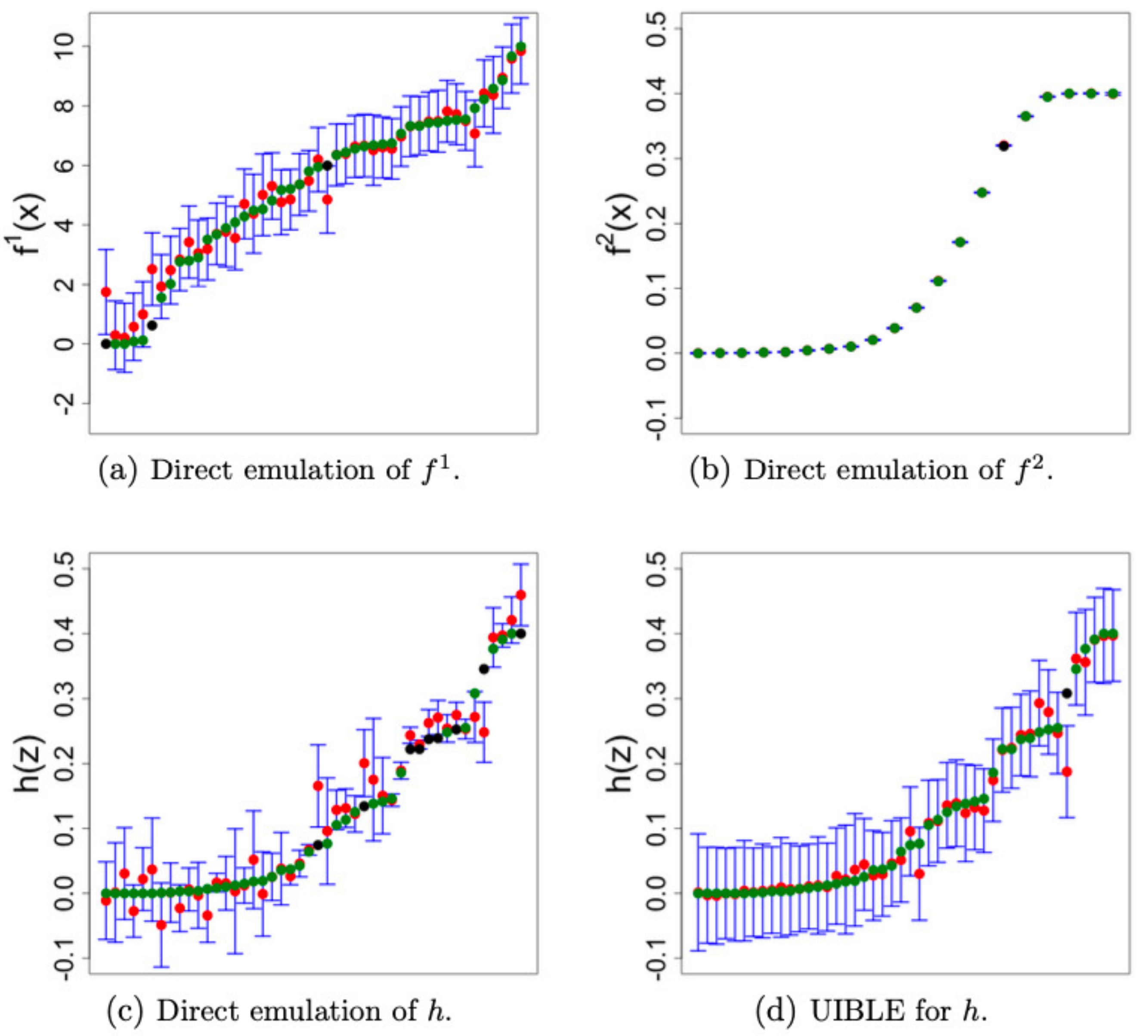}
\caption{Adjusted expectation (red points) $\pm3$ standard deviations for each of the diagnostic points (indexed along the x-axis according to increasing size of the simulator output) when 50 simulator training points were used for $f^1$, and $h$, and 20 were used for $f^2$, with the simulator output value given as a green point if falling within the $\pm3$ standard deviation error bar, and black otherwise.}
\label{DDR_diagnostics_50}
\end{figure}

\begin{figure}
\centering
\includegraphics[height=11.5cm, width = 12.3cm, angle = 0]{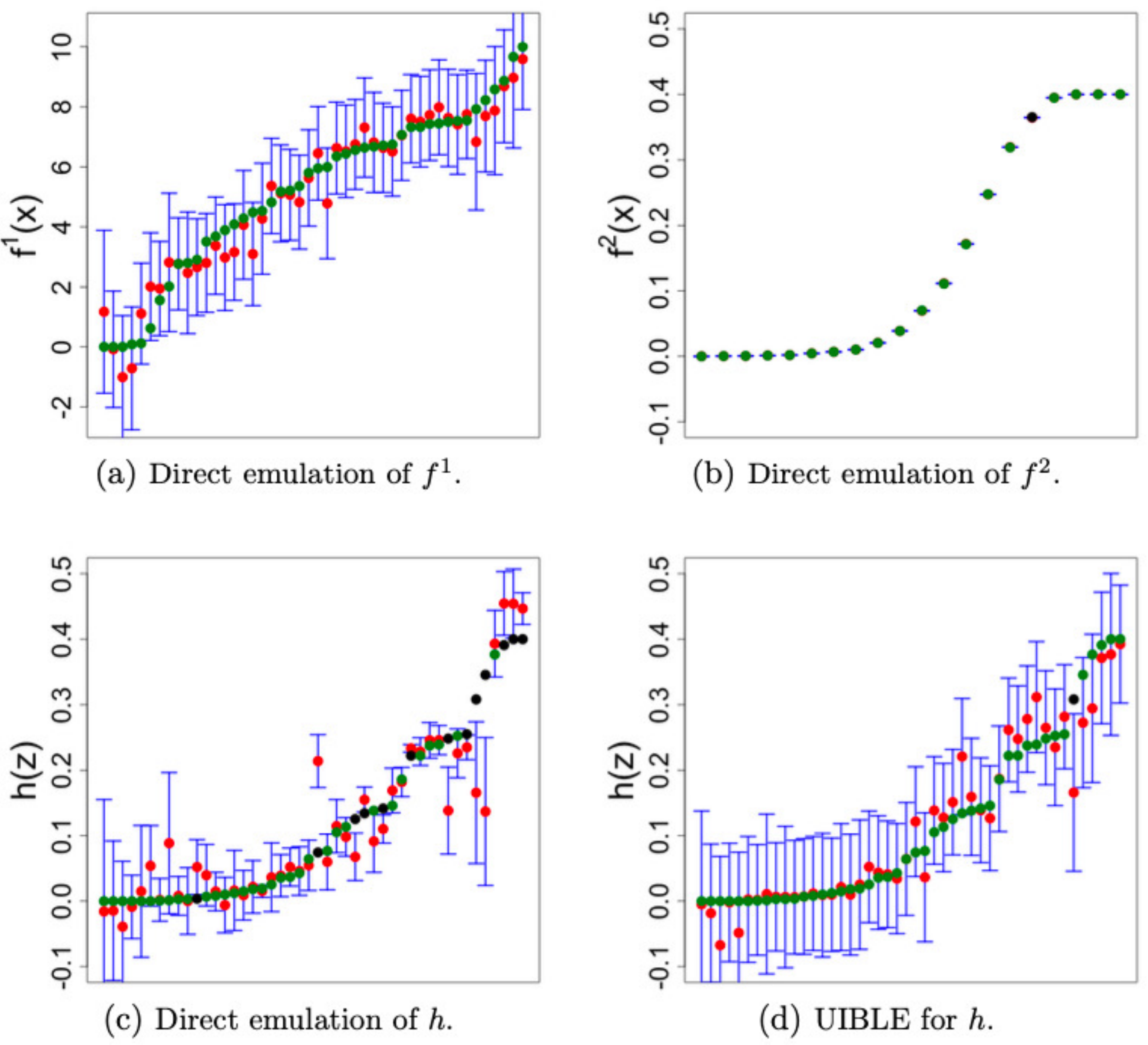}
\caption{Adjusted expectation (red points) $\pm3$ standard deviations for each of the diagnostic points (indexed along the x-axis according to increasing size of the simulator output) when 30 simulator training points were used for $f^1$, $f^2$ and $h$, with the simulator output value given as a green point if falling within the $\pm3$ standard deviation error bar, and black otherwise.}
\label{DDR_diagnostics_30}
\end{figure}

Given the component emulators for $ f^1 $ and $ f^2 $, we can then combine them using UIBLE to yield chained emulators for $ h $.  
Figures \ref{DDR_diagnostics_50} and \ref{DDR_diagnostics_30} show plots, for the case of 50 and 30 training points respectively, of  adjusted expectation (red points) $\pm3$ standard deviations for each of a set of diagnostic points (indexed along the x-axis according to increasing size of the simulator output), with the simulator output value given as a green point if falling within the $\pm3$ standard deviation error bar, and black otherwise.
% $ \pm 3 $ standard deviations against simulator output for a set of diagnostic runs for four different approximations; DE of $ f^1 $, $ f^2 $ and $ h $, then approximation of $ h $ via 
% UIS (using normal and uniform sampling distributions) and 
% UIBLE for $h$.
The input designs for these diagnostic runs were constructed in the same manner as the training run designs.  

In addition to the plots, Table \ref{RMSE_Tab} shows the Mean Absolute Standardised Prediction Error (MASPE):
\be
\frac{1}{n} \sum_{k=1}^n \frac{ | f(\siin^{(k)}) - \mu_f(\siin^{(k)}) | }{\sqrt{\nu_f(\siin^{(k)})}}  ,\label{eq_MASPE}
\ee
Root Mean Squared Prediction Error (RMSPE)  \citep{DGPE}:
\be
\sqrt{ \frac{1}{n} \sum_{k=1}^n (f(\siin^{(k)}) - \mu_f(\siin^{(k)}) )^2 }  \label{eq_RMSE} ,
\ee
and Mean Generalised Entropy Score (MGES), as defined by Equation~(27) of \cite{gneiting2007sps}:
\be
- \, \frac{1}{n}\sum_{k=1}^n \left \{ \frac{ \left[ f(\siin^{(k)}) - \mu_f(\siin^{(k)}) \right] ^2 }{\nu_f(\siin^{(k)})} + \log(\nu_f(\siin^{(k)}))  \right \} \label{eq_GR27}
\ee
for the diagnostic runs for each of the four emulators, with 
% $ \psi = f,g $ or $ h $ as appropriate and 
$ \mu_f $, $ \nu_f $ representing appropriate mean and variance estimators corresponding to generic simulator output $ f $.  MASPE is a measure of emulator validity; heuristically we expect this value to be roughly 1 (assuming normal errors this value should be $ \sqrt{2/\pi} $).  RMSPE permits comparison of emulator accuracy.  MGES is larger (better) for approximations that are both valid and precise.

\begin{table}
\caption{The MASPE, RMSPE and MGES for each of the four approximations and two design sizes discussed in Section~\ref{sec:DDR}. MASPE and RMSPE are smaller-the-better quantities; MGES is larger-the-better. \label{RMSE_Tab}}
  \centering
\fbox{%
\begin{tabular}{c|c|cc|cc}
& & DE of $f^1$ & DE of $f^2 $ & \,\,\, DE \,\,\, & \, UIBLE \, \\ \hline
& MASPE & 0.850 &  0.580 & 1.676  & 0.485    \\
50-50-20 & RMSPE & 0.520 & 0.00031 & 0.0308  & 0.0223 \\
& MGES & 0.343 & 15.161 & 3.960  & 6.616 \\ \hline
& MASPE & 0.798 &  0.987 & 2.075  & 0.754    \\
30-30-30 & RMSPE & 0.616 & 0.00025 & 0.0534  & 0.0379 \\
& MGES & 0.111 & 15.176 & -0.019  & 5.683 \\
\end{tabular}
}
\end{table}

We can see from Figure \ref{DDR_diagnostics_50}a that the emulator for $ f^1 $ is fairly accurate, with less accuracy where simulator output lies towards the lower end of the output range.
% with the exception of points towards the bottom end of the output range, where there are several cases of severe overestimation (with underestimated uncertainty).  
The emulator for $ f^2 $ (Figure \ref{DDR_diagnostics_50}b) is very accurate, reflecting the fact that emulator predictions can be taken with almost as much certainty as running the simulator itself. 
% As a result, this example serves also to demonstrate the applicability of our methods of approximating simulator networks when only some of the simulators are computationally intensive enough to warrant emulating.
% The diagnostics for these emulators should be kept in mind as we discuss the emulators for $ h $.
The direct emulator for $ h $ (Figure \ref{DDR_diagnostics_50}c) yields predictions with underestimated uncertainty.  
By comparison, the estimated uncertainty for UIBLE is perhaps slightly overestimated, but in general yields both more appropriate MASPE values and improved MGES values.  
%By comparison, the estimated uncertainty for the remaining methods is larger, yielding both more appropriate MASPE values and improved MGES values.  
In addition, the accuracy of the predictions for the UIBLE are, on the whole, improved, this being confirmed by the RMSPE values.
% in Table \ref{RMSE_Tab}.  The RMSPE values 
% for UIBLE.  
% It is interesting to note, however, that the uncertainty attributed to each diagnostic point is different between the two approximations, with 
%The uncertainty of UIBLE is larger for runs resulting in low or high values of $ h(\cin) $, and smaller for those points in the middle.  
% This is likely to be a consequence of the way the uncertainty in $ f^1 $ is propagated through $ f^2 $ in the two methods.  
% \sam{UIS} propagates uncertainty in $ f^1 $ by sampling possible values of $ f^2 $ according to possible values of $ f^1 $.  
% This results in a heteroscedastic error structure across the emulator for $ h $ (for example, if $ f^2 $ is expected to change little regardless of the possible values of $ f^1 $, the uncertainty is small).  

Looking now to Figure \ref{DDR_diagnostics_30}, we can see greater estimated uncertainty and lower levels of accuracy for the direct emulators of $f^1$ and $h$, as well as the UIBLE, relative to the corresponding predictions made using the emulators trained with 50 simulator runs.
What is consistent for both design sizes, however, is that direct emulation underestimates the uncertainty towards the upper end of the input space, where the predictions are in addition more inaccurate than those resulting from UIBLE.

It is particularly noticable, in both Figures \ref{DDR_diagnostics_50}d and \ref{DDR_diagnostics_30}d, that UIBLE seems to overestimate uncertainty, particularly where the simulator output lies towards the lower end of the output range.  
%In contrast, 
This is likely to be a consequence of the uncertainty propagation approach. 
UIBLE has uncertainty from the regression part and covariance structure. As with standard Bayes linear emulation that uses a single correlation structure across $ \mathbb{X} $ 
% (or in this case $ \expectatio $ and $ \variance [\mathcal{X}] $)
, there is  some averaging of the uncertainty estimates for simulator prediction across the input space, even if the behaviour at some points is smoother than others.  
% Incorporation of more sophisticated methodology into the UIBLE methodology, for example, utilising similar ideas to local GPs \citep{gramacy2015lgp}, may be of benefit in this case.  
Use of uncertain input adaptations to non-stationary covariance structures could address this, and is an area for future research.

To summarise, we feel that the results presented give evidence for utilising UIBLE for linking emulators of component simulators in a network over using a direct emulator of the composite simulator in many cases.  

%%%%%%%%%%%%%%%%%%%%%%%%%%%%%%%%%%%%%%%%%%%%%%

\section{Conclusion \label{conclusion}}

We have presented novel general Bayes linear methodology for the analysis of statistical models with uncertain inputs.
This has been achieved by extending commonly-employed second-order modelling assumptions to be applicable in the context of uncertain inputs.

Section \ref{subsec:EE} demonstrated the applicability of the methodology within the context of a regression model with correlated error structure in the case of known inputs.
By considering adaptations of the constituent components of the prior correlation structure, we developed an appropriate prior covariance structure which could be adjusted under observations (at uncertain inputs) of the modelled electrolysis extraction process over time.
An important feature of the presented adapted covariance structures is the extended homogeneity of the error structure across possible specifications of the set $\{ \e{\bX}, \e{\bX'}, \cov{\bX}{\bX'} \}$.  This homogeneity may result in conservative approximations in the sense of effectively overestimating variances $\var{\by(\bX)}$, and underestimating covariances $\cov{\by(\bX)}{\by(\bX')}$ for $\bX \neq \bX'$, relative to the expected values of the corresponding covariances, that is $\ed{\bX, \bX'}{\cov{\by(\bX)}{\by(\bX')}}$, were we to assume a probability distribution over $\bX$ and $\bX'$ and integrate out.
Such approximations can enter either by conservatively optimising over introduced parameters (such as $p$ in Equation \eqref{eq:eTm} whilst establishing $\cov{J(T)}{J(T')}$), or by making direct statements via inequalities (such as in \eqref{eq:cov_hT} whilst establishing $\cov{H(T)}{H(T')}$).
Whilst practically making little difference, it is perhaps important to bear in mind that whilst the former approximation resolves to the expression for $\cov{J(t)}{J(t')}$ in the case that $\var{T} = \var{T'} = 0$, the latter approximation does not have such resolution.

We justify these covariance structures as useful when we are unwilling or unable to meaningfully specify a probability distribution over the inputs $\bX$ and $\bX'$.
Whilst perhaps deemed conservative, there are many statistical models for which it can be viewed as preferable to underestimate resolved uncertainty in a Bayes linear paradigm than to overestimate resolved uncertainty by arbitrary specification of too concentrated a probability distribution over the inputs, leading to an underestimation of adjusted uncertainty for future inputs, perhaps leading to real-world decision-making with dire consequences.
Having said this, less conservative covariance structures may be available via exploration of heterogeneous covariance structures over the space of possible specifications $\{ \e{\bX}, \e{\bX'}, \cov{\bX}{\bX'} \}$ for $\bX, \bX'$, this being an area for future research.

In Section \ref{UIBLE}, we developed UIBLE, which is a direct extension of Bayes linear emulation methodology to the case of training emulators with, and predicting for, simulator output from uncertain inputs. 
UIBLE is a computationally efficient approximation of simulator output given a second-order belief specification regarding the input, with each evaluation being akin to a run of a standard Bayes linear emulator.
The modelling assumptions are pragmatic, but the resulting emulator can, and should, be assessed using diagnostic summaries and plots for overall adequacy as anyway necessary for a standard Bayes linear emulator (or indeed any approximating statistical model).

The specific application example which we presented was for component-wise emulation of a chain or network of simulators.
In this case, the input to the emulator for the dose-response simulator was uncertain as a result of it being the emulated output of the dispersion simulator.
Such networks of simulators are common, both explicitly as here, and given the fact that many complex simulators, for example climate simulators, are usually made up of smaller components which arguably form a network.
The potential benefits in many scenarios of linking emulators of component-simulators in contrast to direct emulation of the composite simulator is exemplified through the results of Section \ref{sec:DDR}, along with the results of \cite{CCMLSE} and \cite{ming2021lgp}, albeit with these works focusing on Gaussian process emulation rather than Bayes linear emulation.
In some cases, the links between the simulators in a network, and particularly those between the outputs of an earlier simulator and the inputs of a later simulator, may not be directly compatible.
% even if both components are from the same (e.g. climate) model.
In such cases, these differences are a potential source of model discrepancy, which should be taken into account, and an area for future research, however, we have assumed a direct correspondence within this article to aid the illustration of our methodology.

There are future directions for UIBLE in the context of the emulation of a network of simulators. One example would be the consideration of suitable adaptations to additional covariance structures, such as the Mat\'{e}rn correlation function \citep{MEALSPS, DPGP}, or additional adaptations for correlation functions used for the emulation of stochastic simulators with uncertain inputs.

In this work, the two emulators were constructed independently, before being linked together to form an approximation to the composite simulators. 
However, both design and parameter estimation for all component emulators could be considered simultaneously, and provide much scope for future research. 
The issue regarding design has been partially addressed in \cite{ming2021lgp}, where it is correctly highlighted that ignorance of structural dependence cause unnecessary refinements of component emulators that are insignificant to the global output.
They propose a variance-based adaptive design, which selects at each iteration first the component emulator with greatest average variance contribution to the linked emulator, before selecting the single design point for this emulator based on individual contribution to the linked emulator.
Whilst shown to be effective in comparison to the non-sequential design strategy of \cite{CCMLSE}, it should be noted that the emulator with greatest average variance contribution may not contain the optimal simulator run across all component simulators for reducing the targeted aspects of linked emulator variance, for example, average or maximal variance.

A similar design approach based on predictive means in the Bayes linear paradigm, which may get around this issue, is outlined as follows. 
For each simulator $f^{i}$, locate input $\bx_{i}$ with (near-)optimal predicted composite emulator variance reduction by constructing the relevant component emulator under the assumption that $f^i(\bx_i) = \e{f^i(\bx_i)}$, and hence $\var{f^i(\bx_i)} = 0$, for a large range of possible $\bx_i$.
Exploration over possible $\bx_i$ may involve a large space-filling design, or more likely an efficient optimisation algorithm, for each component emulator.
Once the optimal point for each simulator is found, select the simulator for which the component optimal composite emulator variance reduction is greatest.
% A more sophisticated design procedures would be a valuable asset to the emulation of simulator networks.
Whilst we postulate this design algorithm, we do not apply it here because it is not the main focus of this work, and thorough exploration into this topic, which makes advances on this suggestion and the algorithm of \cite{ming2021lgp}, is a separate research area requiring thorough attention. It should be noted in particular that both these algorithms neglect the computational efficiency of the component simulators, which are likely to vary significantly. A realistic and applicable design algorithm, assuming it to be restricted by overall computational budget, must take this into account.

In addition to situations involving networks of simulators, UIBLEs have more general application.
For example, the inputs of a simulator may be uncertain as a result of wishing to make predictions about the corresponding system under a specific scenario, but not knowing the direct relationship between the real-world quantities representing the scenario and the corresponding parameter values in the model.
Alternatively, UIBLEs would permit efficient sensitivity analyses; several evaluations of an emulator with constant $ \e{\bX} $ and varying $ \var{\bX} $ could quickly provide an indication of the influence of individual inputs on simulator output behaviour.

More generally, the methodology presented in this article may fit smoothly within the context of Bayes linear Bayes graphical models and Bayes linear kinematics \citep{BLKBLBGM}.
In particular, both approaches involve specification and adjustment of our beliefs about uncertain quantities within a Bayes linear structure based on revised, or updated, beliefs about other quantities within the structure.
We view this connection as an area for future research, since the inputs to the statistical model over which we are uncertain may enter into the modelling process in a non-linear way, for example, through use of correlation functions.
In addition, we wish the specification, of model output, to hold for all possible specifications $\{ \e{\bX}, \var{\bX} \}$ of the input variable $\bX$ that we may have.

To conclude, this work opens up many doors for the analysis of statistical modelling with uncertain inputs, particular within a Bayes linear paradigm.
The most appropriate belief structures for different practical applications will be situation-specific, and scope for future research themselves, but will benefit from the foundations laid out within this article.

%%%%%%%%%%%%%%%%%%%%%%%%%%%%%%%%

\subsection*{Acknowledgements}

% This work was undertaken as part of the ``Crystalcast'' project funded by the US Defence Threat Reduction Agency. 
This work was supported by Chemical and Biological Technologies Department (contract HDTRA1-17-C-0028).
We are grateful to Crystalcast project members for invaluable discussions, comments, and provision of the simulators for the dispersion dose-response application. Particular thanks are due to Professor Veronica Bowman and Dr Daniel Silk (Defence Science and Technology Laboratory, UK), and Dr Daria Semochkina (University of Southampton, UK).

%%%%%%%%%%%%%%%%%%%%%%%%%%%%%%%%

\bibliographystyle{agsm}

\bibliography{Biblionotes}

%%%%%%%%%%%%%%%%%%%%%%%%%%%%%%%%%%%%%%%%%%%%%%

\appendix

%%%%%%%%%%%%%%%%%%%%%%%%%%%%%%%%%%%%%%%%%%%%%%

\section{Extended Calculations for Section 3 \label{ExchangeableRegressionsEgDerivations}}

We here present the material covered in Section 3 of the main text again, but with extended detail in calculation for the interested reader.

Consider a regression model, presented here with scalar output $y$ for simplicity of notation and exposition:
 \be
 y(\bX) = \bX^T \bb + \epsilon(\bX) \label{eq:reg_mod}
 \ee
 where $\bb \in \real^p$ is a vector of regression parameters, and $\bX \in \real^p$ is a vector random variable. 
 
 We consider a generic prior specification over $\bb$ of the form $ \e{\bb} = \bGa$ and $ \var{\bb} = \bDe $. 
 Under the usual scenario of known $ \bx $, we are likely to specify $ \e{\epsilon(\bx)} = 0 $, thus we view it a reasonable extension to specify $ \e{\epsilon(\bX)} = 0 $ in the random variable scenario.
 It is common for $\epsilon$ to be viewed as uncorrelated with $\bb$ \emph{a priori}, that is to specify $\cov{\bb}{\epsilon(\bx)} = 0$, so again we view it as reasonable to extend this in the random variable context to $\cov{\bb}{\bX} = 0$.   
 % We must also specify a correlation structure over $\epsilon(\bX)$.  We consider two example cases of such a specification in Sections \ref{subsec:LR} and \ref{subsec:EE}, however, we first go through the calculations that don't require this distinction.
 
Following the above prior specifications, adjusted belief specification for $y(\bX)$ can be obtained as follows:
\ba 
\ed{\my}{y(\bX)} & = & \e{y(\bX)} + \cov{y(\bX), \my} \, \var{\my}^{-1} \, (\my - \e{\my}), \\
\covd{\my}{y(\bX)}{y(\bX')} & = & \cov{y(\bX)}{y(\bX')} - \cov{y(\bX)}{\my} \, \var{\my}^{-1} \, \cov{\my}{y(\bX')},
\ea
where
\ba
\e{\my} & = & (\seq{\e{y(\bX^{(1)})}}{\e{y(\bX^{(n)})}} ), \\ 
\var{\my} & = & \{ \cov{y(\bX^{(i)})}{y(\bX^{(j)})} \}_{i,j = \seq{1}{n}} .
 \ea
 Therefore the required prior specifications are 
 $ \e{y(\bX)} $ and $ \cov{y(\bX)}{y(\bX')} $
% over second-order belief specifications 
for any $ \bX, \bX' \in \mbX $.
 
We have that:
\ba
\e{y(\bX)} & = & \e{\bX^T \bb} + \e{\epsilon(\bX)} 
\NLeq
\e{\bX^T} \, \bGa,
\ea
and
\ba
\cov{y(\bX)}{y(\bX')} & = &  \cov{\bX^T \bb + \epsilon(\bX)}{\bX'^T \bb + \epsilon(\bX')} \nonumber  \\
& = & \cov{\bX^T \bb}{\bX'^T \bb} + \cov{\epsilon(\bX)}{\epsilon(\bX')} \nonumber  \\
& & \,\, + \, \cov{\bX^T \bb}{\epsilon(\bX')} + \cov{\epsilon(\bX)}{\bX'^T \bb}
\NLeq
 \cov{\bX^T \bb}{\bX'^T \bb} + \cov{\epsilon(\bX)}{\epsilon(\bX')},  \label{eq:cov_yX}
\ea
with the final line of Equation \eqref{eq:cov_yX} resulting from the law of total covariance:
\ba
\lefteqn{\cov{\bX^T \bb}{\epsilon(\bX')}}
\NLeq
\e{\cov{\bX^T \bb}{\epsilon(\bX') | \bX = \bx, \bX' = \bx' } } + \cov{\e{\bX^T \bb | \bX = \bx }}{\e{\epsilon(\bX') | \bX' = \bx' }}
\NLeq
\e{\bX^T \, \cov{\bb}{\epsilon(\bX')}} + \cov{\e{\bX^T \bb | \bX = \bx }}{0}
\NLeq
0 + 0 = 0.
\ea
For the first term of Equation~\eqref{eq:cov_yX}, we have, again using the law of total covariance, that
\ba
\lefteqn{\cov{\bX^T \bb}{\bX'^T \bb}} \\
& = & \e{\cov{\bX^T \bb}{\bX'^T \bb | \bX = \bx, \bX' = \bx' } } + \cov{\e{\bX^T \bb | \bX = \bx }}{\e{\bX'^T \bb | \bX' = \bx' }} \nonumber \\
% & = & \e{ \bX^T \bDe \bX' } + \cov{\bX^T \e{\bb}}{\bX'^T \e{\bb}} \nonumber \\
& = & \e{ \bX^T \bDe \bX' } + \bGa^T \cov{\bX}{\bX'} \bGa,
\ea
with
\ba
\e{ \bX^T \bDe \bX' } & = & \e{ \trace(\bX^T \bDe \bX') } 
\NLeq
 \e{ \trace(\bDe \bX' \bX^T) } 
 \NLeq
 \trace( \e{\bDe \bX' \bX^T} )
 \NLeq
 \trace( \bDe \e{ \bX' \bX^T } )
 \NLeq
 \trace( \bDe( \e{\bX'} \e{\bX^T} + \cov{\bX'}{\bX} ) ).
\ea

 % The regression model can be viewed as exchangeable over different $\bb$.
%Following the above prior specifications, adjusted belief specification for $y(\bX)$ can be obtained as follows:
%\ba 
%\ed{\my}{y(\bX)} & = & \ed{\my}{\bX^T \bb} + \ed{\my}{\epsilon(\bX)} \\
%%
%\covd{\my}{y(\bX)}{y(\bX')} & = & \covd{\my}{\bX^T \bb + \epsilon(\bX)}{\bX'^T \bb + \epsilon(\bX')} \nonumber  \\
%%
%& = & \covd{\my}{\bX^T \bb}{\bX'^T \bb} + \covd{\my}{\epsilon(\bX)}{\epsilon(\bX')} \nonumber  \\
%%
%& & \,\, + \covd{\my}{\bX^T \bb}{\epsilon(\bX')} + \covd{\my}{\epsilon(\bX)}{\bX'^T \bb}
%\ea
%
%Making the specification that $\bX$ is independent (\sam{uncorrelated?}) of $\my$, we have that
%\ba
%\ed{\my}{\bX^T \bb} & = & \ed{\my}{\bX^T} \ed{\my}{\bb} \nonumber \\
%%
%& = & \e{\bX^T} \ed{\my}{\bb} 
%\ea
%where, using the Bayes Linear Update Equations (BLUEs)
%\be
%\ed{\my}{\bb} = \bGa + \cov{\bb}{\my} \, \var{\my}^{-1} \, (\my - \e{\my}) 
%\ee
%with
%\be
%\cov{\bb}{\my} = \cov{\bb}{\mx \bb + \epsilon(\mx)} = \bDe \, \mx^T
%\ee
%and 
%\ba
%\e{\my} & = & \e{\mx^T \bb} = \e{\mx^T} \bGa \nonumber \\ 
%\var{\my} & = & \{ \cov{y(\bX^{(i)})}{y(\bX^{(j)})} \}_{i,j = \seq{1}{n}}
%\ea
%with
%\ba
%\cov{y(\bX)}{y(\bX')} & = &  
%\ea
%
%We also have
%\be
%\ed{\my}{\epsilon(\bX)} = \cov{\epsilon(\bX)}{\my} \, \var{\my}^{-1} \, (\my - \e{\my}) 
%\ee
%and
%\be
%\covd{\my}{\bX^T \bb}{\bX'^T \bb} = \cov{\bX^T \bb}{\bX'^T \bb}  - \cov{\bX^T \bb}{\my} \, \var{\my}^{-1} \, \cov{\my}{\bX'^T \bb} 
%\ee

The second term of Equation~\eqref{eq:cov_yX} is a covariance specification over the residual process $\epsilon(\bX) $.  
In Sections \ref{subsec:LR} and \ref{subsec:EE}, we present two example model belief structures in the context of statistical modelling with known inputs, demonstrating reasonable generalisation to the random variable input scenario using the ideas presented in Section 2 of the main text.

%%%%%%%%%%%%%%%%%%%%%%%%%%%%

\subsection{Linear Regression with Uncorrelated Random Error \label{subsec:LR}}

In the common known input case, the standard linear regression model with uncorrelated and homoscedastic random error can be obtained by specifying a prior covariance structure over $ \epsilon(\bx) $ of
\be
\cov{\epsilon(\bx)}{\epsilon(\bx')} = \indicator_{\bx = \bx'} \sigma^2,
\ee
where $ \indicator $ is the indicator function taking value $1$ if the statement is true and $0$ otherwise, and $\sigma^2$ is a common scalar variance parameter.

This specification can be simply extended to the random variable case as follows:
\be
\cov{\epsilon(\bX)}{\epsilon(\bX')} = \indicator_{\bX = \bX'} \sigma^2,
\ee
where the covariance across two random variable inputs is zero unless they are known to be the same.  Note that we could have two non-identical random variables with the identical second order belief specification.

%%%%%%%%%%%%%%%%%%%%%%%%%%%%

\subsection{Example: Extracting Aluminium by Electrolysis\label{subsec:EE}}

\sam{State that $t\geq3$ in order to generalise to continuous $t$ values.}

In this section, we consider an adaptation of the exchangeable regressions example for extracting aluminium by electrolysis over time presented in \cite{goldstein1998aeb}.
We first consider the covariance structure as presented in that paper with known input, before presenting a reasonable generalisation to the random variable input scenario using the ideas presented in Section 2 of the main text.

The model is of the form 
\be
y(t) = \beta_0 +  \beta_1 \, t + \epsilon(t),
\ee
and is exchangeable over $\bb$.
Note that this regression model is consistent with the general form presented in Equation~\eqref{eq:reg_mod}, with $p=2$ variables being an intercept and single controllable parameter $t$.
In \cite{goldstein1998aeb}, it is assumed that $t=\seq{1}{13}$, as well as a structured error model as follows:
\ba
\epsilon(t) & = & A(t) + J(t) + H(t),
\\ \nonumber
%%
%A(t) & = & \indicator_{t = t'} \sigma^2
%\\ \nonumber
%
J(t) & = & J(t-1) + Q(t),
\\ \nonumber
H(t) & = & \psi \, H(t-1) + R(t),
\ea
with $ \psi \in (0,1) $, and where, for generality, we denote
$ \var{A(t)} = \sigma^2_A $,
$ \var{Q(t)} = \var{J(1)} = \sigma^2_Q $,
$ \var{R(t)} = \sigma^2_R $, 
and
$ \var{H(1)} = \sigma^2_1 $.
These terms express discrepancies from the linear trend as the sum of a pure measurement error $A(t)$, a stochastic development of the discrepancy as a random walk with drift $J(t)$, and an autoregressive term expressing the measurement of the suspended particles in the chemical analysis $H(t)$.  For further details see \cite{goldstein1998aeb}.  Whilst the original model assumes discrete timesteps $t$, we will generalise the regression structure to continuous $t$-values, subject to $t \geq 3$ for consistency issues, before addressing the uncertain input $T$ scenario.

In this example, we have that:
\ba
\cov{\epsilon(t)}{\epsilon(t)} & = &
\cov{A(t) + J(t) + H(t)}{A(t') + J(t') + H(t')}
\NLeq
\cov{A(t)}{A(t')} + \cov{J(t)}{J(t')} + \cov{H(t)}{H(t')}.  \label{eq:cov_AJH}
\ea

We consider each term of Equation~\eqref{eq:cov_AJH} separately.
To begin with, we have
\be
\cov{A(t)}{A(t')} = \indicator_{t=t'} \sigma_A^2. \label{eq:cov_A}
\ee
Secondly, defining $d = |t-t'|$, and $t_m = \min(t,t')$, we have
\ba
\cov{J(t)}{J(t')} & =  & \cov{J(t_m)}{J(t_m+d)}
\NLeq
\cov{J(t_m)}{J(t_m) + Q_{t_m+1} + ... + Q_{t_m+d}}
\NLeq
\var{J(t_m)}
\NLeq
t_m\,\sigma^2_Q,   \label{eq:cov_J}
\ea
resulting in a symmetric correlation structure as follows:
\ba
\corr{J(t)}{J(t')} & = & \frac{\cov{J(t)}{J(t')}}{\sqrt{\var{J(t)}\var{J(t')}}}
\NLeq
\frac{\cov{J(t_m)}{J(t_m+d)}}{\sqrt{\var{J(t_m)}\var{J(t_m+d)}}}
\NLeq
\frac{t_m \, \sigma^2_Q}{\sqrt{t_m \,  \sigma^2_Q \, (t_m+d) \, \sigma^2_Q }}
\NLeq
\sqrt{\frac{t_m}{t_m+d}}.  \label{eq:corr_J}
\ea

Similarly,
\ba
\cov{H(t)}{H(t')} & =  & \cov{H(t_m)}{H(t_m+d)}
\NLeq
\cov{H(t_m)}{\psi^d\,H(t_m) + \psi^{d-1}\,R_{t_m+1} + ... + \psi\,R_{t_m+d}}
\NLeq
\psi^d\,\var{H(t_m)}
\NLeq
\psi^d \, \left( \psi^{2(t_m-1)} \, \sigma^2_1 + \frac{1 - \psi^{2(t_m-2)}}{1 - \psi^2} \, \sigma_R^2 \right),  \label{eq:cov_H}
\ea
since 
\ba
\var{H(t)} & = & \var{\psi^{t-1}\,H_1 + \psi^{t-2}\,R_2 + ... +  \psi \, R_{t-1} + R_{t}}
\NLeq
\psi^{2(t-1)} \, \sigma^2_1 \, + \, \psi^{2(t-2)} \, \var{R_2} + ... + \psi^2 \, \var{R_{t-1}} + \var{R_{t}}
\NLeq
\psi^{2(t-1)} \, \sigma^2_1 + ( \psi^{2(t-2)} + \psi^{2(t-3)} + ... + \psi^2 + 1 ) \, \sigma_R^2
\NLeq
\psi^{2(t-1)} \, \sigma^2_1 + \frac{1 - \psi^{2(t-2)}}{1 - \psi^2} \, \sigma_R^2,  \label{eq:var_H}
\ea
resulting in a symmetric correlation structure as follows:
\ba
\corr{H(t)}{H(t')} & = & \frac{\cov{H(t)}{H(t')}}{\sqrt{\var{H(t)}\var{H(t')}}}
\NLeq
\frac{\cov{H(t_m)}{H(t_m+d)}}{\sqrt{\var{H(t_m)}\var{H(t_m+d)}}}
\NLeq
\frac{\psi^d\,\var{H(t_m)}}{\sqrt{\var{H(t_m)}\var{H(t_m+d)}}}
\NLeq
\psi^d \, \sqrt{\frac{\var{H(t_m)}}{\var{H(t_m+d)}}}
\NLeq
\psi^d \, \sqrt{\frac{\psi^{2(t_m-1)} \, \sigma^2_1 + \frac{1 - \psi^{2(t_m-2)}}{1 - \psi^2} \, \sigma_R^2}{\psi^{2(t_m+d-1)} \, \sigma^2_1 + \frac{1 - \psi^{2(t_m+d-2)}}{1 - \psi^2} \, \sigma_R^2}}
\NLeq
\psi^d \, \sqrt{\frac{(1 - \psi^2) \psi^{2(t_m-1)} \, \sigma^2_1 + (1 - \psi^{2(t_m-2)}) \, \sigma_R^2}{(1 - \psi^2) \psi^{2(t_m+d-1)} \, \sigma^2_1 + ( 1 - \psi^{2(t_m+d-2)})  \, \sigma_R^2}}.  \label{eq:corr_H}
\ea

Combining the results of Equations \eqref{eq:cov_A}, \eqref{eq:cov_J} and \eqref{eq:cov_H} with Equation \eqref{eq:cov_AJH} leads to a prior belief structure across all $t$ of
\ba
\cov{\epsilon(t)}{\epsilon(t)} & = & \indicator_{t=t'} \sigma_A^2 + t_m\,\sigma^2_Q  + \psi^d \, \left( \psi^{2(t_m-1)} \, \sigma^2_1 + \frac{1 - \psi^{2(t_m-2)}}{1 - \psi^2} \, \sigma_R^2 \right). 
\ea 

We now proceed to consider suitable extensions to the uncertain input scenario.  In this case, we need an expectation and covariance structure spanning $\e{T}, \e{T'}, \cov{T}{T'}$-space.  
We consider that the statements of expectation in the error structure simply extend to being $\e{A(T)} = 0$, $\e{J(T)} = 0$ and $\e{H(T)} = 0$, with logical derivation of these statements being shown using the law of total expectation.
Regarding covariance structure, Equation \eqref{eq:cov_A} simply extends to the following:
\be
\cov{A(T)}{A(T')} = \indicator_{T=T'} \sigma_A^2, \label{eq:covUI_A}
\ee
this being similar to the uncorrelated error term shown in Section \ref{subsec:LR} of this supplementary material.

Equation \eqref{eq:cov_J} extends as follows, where the random variables $T_m = \min(T,T')$ and $D=|T-T'|$ correspond to known quantities $t_m$ and $d$ above:
\ba
\lefteqn{\cov{J(T)}{J(T')}} 
\NLeq
\cov{J(T_m)}{J(T_m+D)}
\NLeq
\e{\cov{J(T_m)}{J(T_m+D) | T_m=t_m, D=d }} + \cov{\e{J(T_m)|T_m}}{\e{J(T_m+D)|T_m,D}}
\NLeq
\e{T_m \, \sigma^2_Q} + 0
\NLeq
\sigma^2_Q \, \e{T_m}.
\ea
The tricky specification here is eliciting $ \e{T_m} $ from $ \e{T}, \e{T'} $ and $ \cov{T}{T'} $, since a function which takes the minimum of two random quantities is a non-linear function.  

If we know which of our random variables is bigger (without loss of generality, that $T < T'$, say), then $\e{T_m} = \e{T}$.  This is not an unrealistic situation; for example, we may have measurements at two times about which we are uncertain, but know that one of them certainly happened after the other.
% Perhaps we specify \cov{T}{T'} to be commensurate with this, or perhaps it doesn't need commenting on as it is no longer relevant.

More generally, we propose a reasonable covariance structure as follows.
%which satisfies several requirements, including;
% \bi
% \item The expected value of the covariance function assuming the conditional covariance structure as for the known input case should be an upper bound on the proposed covariance model specification.
% \item The proposed covariance structure should resolve to the covariance structure for the known input case when the inputs are indeed known (that is, when $\e{T} = t$, $\e{T'} = t'$ and $ \cov{T}{T'} = 0 $).
% \ei
Let $ \ind = \indicator_{T<T'} $ and $p = \e{\ind}$. We then condition on this event using the law of total expectation to give:
\ba
\e{T_m} & = & \e{ \ed{\ind}{T_m}}
\NLeq
\ed{\ind = 1}{T_m} \, \e{\ind} + \ed{\ind = 0}{T_m} \,( 1 - \e{\ind} )
\NLeq
p \, \ed{\ind = 1}{T_m} + ( 1 - p ) \, \ed{\ind = 0}{T_m}, \label{eq:Tm_Total_Exp}
\ea
where
\ba
\ed{\ind = 1}{T_m} & = & \ed{\ind = 1}{T}
\NLeq
\e{T} + \cov{T}{\ind} \, \var{\ind}^{-1} \, (\ind - \e{\ind} )
\NLeq
\e{T} + \frac{\cov{T}{\ind}}{p(1-p)} \times (1-p) \,
\NLeq
\e{T} + \frac{\cov{T}{\ind}}{p},
\\
\ed{\ind = 0}{T_m} & = & \ed{\ind = 0}{T'}
\NLeq
\e{T'} + \cov{T'}{\ind} \, \var{\ind}^{-1} \, (\ind - \e{\ind} )
\NLeq
\e{T'} + \frac{\cov{T'}{\ind}}{p(1-p)} \times (-p) \,
\NLeq
\e{T'} - \frac{\cov{T'}{\ind}}{1-p},
\ea
and
\ba
\cov{T}{\ind} & = & \corr{T}{\ind} \, \sqrt{\var{T} \, \var{\ind} }
\NLeq
\corr{T}{\ind} \, \sqrt{\var{T} \, p(1-p)},
\\
\cov{T'}{\ind} & = & \corr{T'}{\ind} \, \sqrt{\var{T'} \, \var{\ind} }
\NLeq
\corr{T'}{\ind} \, \sqrt{\var{T'} \, p(1-p)},
\ea
so that:
\ba
\e{T_m} & = & p \left( \e{T} + \frac{1}{p} \, \corr{T}{\ind} \, \sqrt{\var{T} \, p(1-p)} \right)
\NL
+ \, (1-p) \left( \e{T'} - \frac{1}{1-p} \, \corr{T'}{\ind} \, \sqrt{\var{T'} \, p(1-p)} \right)
\NLeq
p \, \e{T} + (1-p) \, \e{T'}
\NL
+ \sqrt{p(1-p)} \left( \corr{T}{\ind} \, \sqrt{\var{T}} - \corr{T'}{\ind} \, \sqrt{\var{T'}} \right).   \label{eq:eTm}
\ea

We may be able to specify values for all quantities required in Equation \eqref{eq:eTm}, in which case we have our expression for $\e{T_m}$.  If specification of $\corr{T}{\ind}$ and $\corr{T'}{\ind}$ is proving difficult, then let us first note that it is logical to assume that 
$ -1 < \corr{T}{\ind} < 0 $
and
$ 0 < \corr{T'}{\ind} < 1 $, so that
\be
\e{T_m} > p \, \e{T} + (1-p) \, \e{T'} - \sqrt{p(1-p)} \left( \sqrt{\var{T}} + \sqrt{\var{T'}}  \right). \label{eq:eTm_bound}
\ee

Again, if we feel able to specify a value for $p$, then we can use it in Equation \eqref{eq:eTm_bound}. 
Alternatively, in order to be conservative with respect to variance resolution, we can minimise the exspression on the right had side of Equation \eqref{eq:eTm_bound} over $p \in [0,1]$.

First, we note that both Equations~\eqref{eq:eTm} and \eqref{eq:eTm_bound} are of the form
\be
h(p) = ap + b(1-p) - c \sqrt{p(1-p)}, \label{eq:hp}
\ee
with 
$ a = \e{T} $,
$ b = \e{T'} $,
and 
$ c = \sqrt{\var{T}} + \sqrt{\var{T'}} $
or
$ c = \corr{T}{\ind} \, \sqrt{\var{T}} - \corr{T'}{\ind} \, \sqrt{\var{T'}} $.

Use elementary calculus to find the minimum, by first taking the derivative of $h(p)$ with respect to $p$:
\be
\frac{\mathrm{d} h(p)}{\mathrm{d} p} = a - b + \frac{c(2p-1)}{2\sqrt{p(1-p)}}.
\ee
We then set 
$ \frac{\mathrm{d} h(p)}{\mathrm{d}p} = 0 $
to find $ p $ which minimises $h(p)$:
\ba
&& a - b + \frac{c(2p-1)}{2\sqrt{p(1-p)}} = 0
\nonumber \\ & \implies &
%
% \frac{c(2p-1)}{2\sqrt{p(1-p)}} = b - a
% \nonumber \\ & \implies &
% %
% \frac{c^2(2p-1)^2}{4p(1-p)} = (b - a)^2
% \nonumber \\ & \implies &
%
c^2(2p-1)^2 = 4p(1-p)(b-a)^2
\nonumber \\ & \implies &
%
% c^2(4p^2 - 4p + 1) = 4(p-p^2)(b-a)^2
% \nonumber \\ & \implies &
%
p^2(4c^2 + 4(b-a)^2) - p(4c^2 + 4(b-a)^2) + c^2 = 0,
\ea
which is of the form
\be
kp^2 - kp + c^2 = 0, \label{eq:kp}
\ee
with 
$ k = 4(c^2 + (b-a)^2) $.
Equation \eqref{eq:kp} has the solution
\ba
\hat{p}  & = & % \frac{k \pm \sqrt{k^2 - 4kc^2}}{2k}
% \NLeq
% %
% \half \left( 1 + \sqrt{ \frac{k^2 - 4kc^2}{k^2} }  \right)
% \NLeq
% %
% \half \left( 1 + \sqrt{ \frac{k - 4c^2}{k} }  \right)
% \NLeq
% %
% \half \left( 1 + \sqrt{ \frac{4(c^2 + (b-a)^2) - 4c^2}{4(c^2 + (b-a)^2)} }  \right)
% \NLeq
% %
% \half \left( 1 + \sqrt{ \frac{(b-a)^2}{c^2 + (b-a)^2} } \right) 
% \NLeq
% %
\half \left( 1 + \sqrt{ \mn } \right), 
\ea
where
$ m = b - a $
and
$ n = c^2 + m^2 $.

Inserting this into Expression~\eqref{eq:hp}, we have
\ba
h(\hat{p}) & = & a \hat{p} + b (1 - \hat{p}) - c \sqrt{\hat{p}(1-\hat{p})}.
\NLeq
% %
% \half (a+b) + (\hat{p} - \half)(a-b) - c \sqrt{\hat{p}(1-\hat{p})}
% \NLeq
% %
% \half (a+b) \pm \half \sqrt{\mn} (a-b) - c \sqrt{ \half \left( 1\pm\sqrt{\mn} \right) \left( 1 - \half \left( 1 \pm \sqrt{\mn} \right) \right) }
% \NLeq
% %
% \half (a+b) \mp \half \sqrt{n} \mn - c \sqrt{ \left( \half \pm \sqrt{\frac{m^2}{4n}} \right) \left( \half \mp \sqrt{\frac{m^2}{4n}} \right) }
% \NLeq
% %
% \half(a + b) \mp \half \sqrt{n} \mn - c \sqrt{\frac{1}{4} - \frac{m^2}{4n}}
% \NLeq
% %
% \half \left( a + b \mp \sqrt{n} \mn - c \sqrt{1 - \mn} \right)
% \NLeq
% %
% \half \left( a + b \mp \sqrt{n} \mn - c \sqrt{\frac{n-m^2}{n}} \right)
% \NLeq
% %
\half \left( a + b - \sqrt{n} \left( \frac{c^2}{n} \pm \mn \right) \right) \label{eq:min}
\ea
The minimum of Expression \eqref{eq:min} occurs when $\pm$ is $+$, hence
\ba
h(\hat{p}) & = & \half \left( a + b - \sqrt{n} \left( \frac{n-m^2}{n} + \mn \right) \right)
\NLeq
% %
% \half \left( a + b - \sqrt{n} \left( \frac{n}{n} \right) \right)
% \NLeq
% %
% \half \left( a + b - \sqrt{n} \right)
% \NLeq
%
\half \left( a + b - \sqrt{(b - a)^2 + c^2} \right). \label{eq:hp_min_max}
\ea
Therefore we have that
\be
\e{T_m} \geq \half \left( \e{T} + \e{T'} - \sqrt{ \left( \e{T'} - \e{T} \right)^2 + \left (\sqrt{\var{T}} + \sqrt{\var{T'}} \right)^2} \right), \label{eq:Tm_lower_bound}
\ee
and we might set
\be
\cov{J(T)}{J(T')} =   \frac{\sigma^2_Q}{2} \left( \e{T} + \e{T'} - \sqrt{ \left( \e{T'} - \e{T} \right)^2 + \left (\sqrt{\var{T}} + \sqrt{\var{T'}} \right)^2} \right).
\ee
From Equation \eqref{eq:cov_J} with $t=t'$, it follows that 
\ba
\var{J(T)} & = & \e{\var{J(T)|T=t}} + \var{\e{J(T)|T}}
\NLeq
\e{\sigma_Q^2 \, T} + 0
\NLeq
\sigma^2_Q \, \e{T},
\ea
since 
$T_m = T$.
We thus make use of an indicator function in the expression for
$ \cov{J(T)}{J(T')} $
to ensure it is appropriate when it is known that $T=T'$: 
\ba
\lefteqn{\cov{J(T)}{J(T')}}
\\ & = &
\frac{\sigma^2_Q}{2} \left( \e{T} + \e{T'} - \sqrt{ \left( \e{T'} - \e{T} \right)^2 + \left( 1 - \indicator_{T=T'} \right) \left (\sqrt{\var{T}} + \sqrt{\var{T'}} \right)^2} \right). \nonumber
\ea

Note that:
\bn 
\item When 
$\var{T} = \var{T'} = 0$, 
the expression on the right hand side of Inequality~\eqref{eq:Tm_lower_bound} yields
\be
\half \left( t + t' - \sqrt{(t-t')^2} \right) = \half \left( t + t' - |t - t'| \right) = \min(t,t'), \nonumber
\ee
thus resolving to the known input scenario.
\item If it is known that $T<T'$, then, as mentioned earlier, 
% and by looking at Equation~\ref{eq:eTm} with $p = 1$, 
it would be logical to have that
$\e{T_m} = \e{T}$.
% This clearly satisfies Inequality~\eqref{eq:Tm_lower_bound}, 
Since
\be
\e{T} = \half \left( \e{T} + \e{T'} - \sqrt{ \left( \e{T'} - \e{T} \right)^2 } \right),
\ee
it is clear that $\e{T_m}$ satisfies Inequality \eqref{eq:Tm_lower_bound} in this case.
% \item We need 
% \be
% \var{J(T)} = \sigma^2_Q \, \e{T}
% \ee
% since 
% \ba
% \var{J(T)} & = & \e{\var{J(T)|T=t}} + \var{\e{J(T)|T}}
% \NLeq
% %
% \e{\sigma_Q^2 \, T} + 0
% \NLeq
% %
% \sigma^2_Q \, \e{T}
% \ea
% This doesn't directly follow from the expression on the right hand side of Inequality~\eqref{eq:Tm_lower_bound}.  
% To remedy this, we can make use of an indicator nugget term.
% For example
% \ba
% \lefteqn{\cov{J(T)}{J(T')}}
% \NLeq
% %
% \frac{\sigma^2_Q}{2} \left( \e{T} + \e{T'} - \sqrt{ \left( \e{T'} - \e{T} \right)^2 + \left( 1 - \indicator_{T=T'} \right) \left (\sqrt{\var{T}} + \sqrt{\var{T'}} \right)^2} \right) \nonumber
% \ea
\en

% Note that:
% \bn 
% \item When 
% $\var{T} = \var{T'} = 0$, 
% the expression on the right hand side of inequality~\eqref{eq:Tm_lower_bound} is just
% \be
% \half \left( t + t' - \sqrt{(t-t')^2} \right) = \half \left( t + t' - |t - t'| \right) = \min(t,t') \nonumber
% \ee
% \item If it is known that $T<T'$, then, as mentioned earlier and by looking at Equation~\ref{eq:eTm} with $p = 1$, it would be logical to have that
% $\e{T_m} = \e{T}$.
% This clearly satisfies Inequality~\eqref{eq:Tm_lower_bound}, since
% \be
% \e{T} = \half \left( \e{T} + \e{T'} - \sqrt{ \left( \e{T'} - \e{T} \right)^2 } \right)
% \ee
% \item We need 
% \be
% \var{J(T)} = \sigma^2_Q \, \e{T}
% \ee
% since 
% \ba
% \var{J(T)} & = & \e{\var{J(T)|T=t}} + \var{\e{J(T)|T}}
% \NLeq
% %
% \e{\sigma_Q^2 \, T} + 0
% \NLeq
% %
% \sigma^2_Q \, \e{T}
% \ea
% This doesn't directly follow from the expression on the right hand side of Inequality~\eqref{eq:Tm_lower_bound}.  
% To remedy this, we can make use of an indicator nugget term.
% For example
% \ba
% \lefteqn{\cov{J(T)}{J(T')}}
% \NLeq
% %
% \frac{\sigma^2_Q}{2} \left( \e{T} + \e{T'} - \sqrt{ \left( \e{T'} - \e{T} \right)^2 + \left( 1 - \indicator_{T=T'} \right) \left (\sqrt{\var{T}} + \sqrt{\var{T'}} \right)^2} \right) \nonumber
% \ea
% \en

We extend Equation \eqref{eq:cov_H} by noticing that
\ba
\cov{H(T)}{H(T')} & = & \cov{H(T_m)}{H(T_m+D)}
\NLeq
\e{\cov{H(T_m)}{H(T_m+D)|T_m=t_m, D = d}} 
\NL
+ \cov{\e{H(T_m)|T_m=t_m}}{\e{H(T_m+D)|T_m=t_m, D = d}}
\NLeq
\e{\psi^D\,\var{H(T_m)}} + 0
\NLeq
\e{\psi^D \, \left( \psi^{2(T_m-1)} \, \sigma^2_1 + \frac{1 - \psi^{2(T_m-2)}}{1 - \psi^2} \, \sigma_R^2 \right)}
\nonumber \\ & \geq &
\e{\psi^D \, \left( \psi^{2(T_m-1)} \, \sigma^2_1 + \sigma_R^2 \right)}
\NLeq
\sigma^2_1 \, \e{\psi^{D+2T_m-2}} + \sigma^2_R \, \e{\psi^D}
\NLeq
\sigma^2_1 \, \exp( \log( \e{\psi^{D+2T_m-2}} )) + \sigma^2_R \, \exp( \log(  \e{\psi^D} ))
\nonumber \\ & \geq &
\sigma^2_1 \, \exp( \e{ \log( \psi^{D+2T_m-2})} ) + \sigma^2_R \, \exp(  \e{\log( \psi^D} ) )
\NLeq
\sigma_1^2 \exp( \log \psi \, \e{D + 2T_m - 2} ) + \sigma_R^2 \exp( \log \psi \, \e{D} )
\NLeq
\sigma_1^2 \exp( \log \psi \, \e{T_M + T_m - 2} ) + \sigma_R^2 \exp( \log \psi \, \e{T_M - T_m} ). \label{eq:cov_hT}
\ea
where the first $\geq$ is a result of the fact that, if 
$T_m\geq3$, 
then 
$(1-\psi^{2(T_m-2)})/(1-\psi^2) \geq 1$,
the second $\geq$ is a result of the fact that $\log$ is a concave function, i.e. that 
$\e{f(X)} \leq f(\e{X})$
if 
$f(X) = \log(X)$.,
and the final line makes use of the fact that 
$ d = t_M - t_m $, where 
$ t_M = \max(t,t') $, and correspondingly
$ D = T_M - T_m$, where
$ T_M = \max(T,T') $.

We note that $ \e{T_M+T_m} = \e{T} + \e{T'}$, 
% (which we should expect and can also be derived by combining Equations~\eqref{eq:Tm_Total_Exp} and \eqref{eq:TM_Total_Exp}), 
and thus seek an upper bound for $\e{T_M - T_m}$ in order to find a lower bound for Expression~\eqref{eq:cov_hT} (since $\log\psi$ is negative if $\psi \in (0,1)$).

Similar to Equation~\eqref{eq:Tm_Total_Exp}, we use the law of total expectation on $\e{T_M}$ to get that
\ba
\e{T_M} & = & \e{ \ed{\ind}{T_M}}
\NLeq
\ed{\ind = 1}{T_M} \, \e{\ind} + \ed{\ind = 0}{T_M} \,( 1 - \e{\ind} )
\NLeq
p \, \ed{\ind = 1}{T_M} + ( 1 - p ) \, \ed{\ind = 0}{T_M}, \label{eq:TM_Total_Exp}
\ea
where:
\ba
\ed{\ind = 1}{T_M} & = & \ed{\ind = 1}{T'}
\NLeq
\e{T'} + \cov{T'}{\ind} \, \var{\ind}^{-1} \, (\ind - \e{\ind} )
\NLeq
\e{T'} + \frac{\cov{T'}{\ind}}{p(1-p)} \times (1-p) \,
\NLeq
\e{T'} + \frac{\cov{T'}{\ind}}{p},
\\
\ed{\ind = 0}{T_M} & = & \ed{\ind = 0}{T}
\NLeq
\e{T} + \cov{T}{\ind} \, \var{\ind}^{-1} \, (\ind - \e{\ind} )
\NLeq
\e{T} + \frac{\cov{T}{\ind}}{p(1-p)} \times (-p) \,
\NLeq
\e{T} - \frac{\cov{T}{\ind}}{1-p}.
\ea

Combining Equations~\eqref{eq:Tm_Total_Exp} and \eqref{eq:TM_Total_Exp}, we get that
\ba
\e{T_M - T_m} & = & \e{T_M} - \e{T_m}
\NLeq
p\, \e{T'} + \cov{T'}{\ind} + (1-p) \, \e{T} - \cov{T}{\ind}
\NL
- \left( p\, \e{T} + \cov{T}{\ind} + (1-p) \, \e{T'} - \cov{T'}{\ind}   \right)
\NLeq
(1-2p) \left( \e{T} - \e{T'} \right) + 2 \left( \cov{T'}{\ind} - \cov{T}{\ind} \right)
\NLeq
(1-2p) \left( \e{T} - \e{T'} \right) 
\NL
+ 2 \left( \corr{T'}{\ind} \sqrt{\var{T'} p(1-p)} - \corr{T}{\ind} \sqrt{\var{T} p(1-p)} \right)
\NLeq
(1-2p) \left( \e{T} - \e{T'} \right) 
\NL
+ 2 \sqrt{p(1-p)} \left( \corr{T'}{\ind} \sqrt{\var{T'}} - \corr{T}{\ind} \sqrt{\var{T}} \right).
\ea

Again, if we don't feel able to specify $\corr{T'}{\ind}$ or $\corr{T}{\ind}$, then using the assumptions that 
$ -1 < \corr{T}{\ind} < 0 $
and
$ 0 < \corr{T'}{\ind} < 1 $,
we have that
\be
\e{T_M - T_m} < (1-2p) \left( \e{T} - \e{T'} \right) 
+ 2 \sqrt{p(1-p)} \left( \sqrt{\var{T'}} + \sqrt{\var{T}} \right). \label{eq:TM_Tm}
\ee

If we have specified a value for $p$, again we can use this in Expression~\eqref{eq:TM_Tm}, otherwise we can maximise over $p\in[0,1]$.  This follows from the calculations to bound $\e{T_m}$ above, since we note again that we have an expression of the form
\be
h(p) = ap + b(1-p) - c \sqrt{p(1-p)}, \label{eq:hp2}
\ee
this time with 
$ a = - \left( \e{T} - \e{T'} \right) = \e{T'} - \e{T} $,
$ b = \e{T} - \e{T'} $,
and 
$ c = -2(\sqrt{\var{T}} + \sqrt{\var{T'}}) $,
which is maximised by Equation~\eqref{eq:min}, this time with $-$ instead of $+$, hence
\ba
h(\hat{p}) & = & \half \left( a + b - \sqrt{n} \left( \frac{c^2}{n} - \mn \right) \right)
\NLeq
% %
% \half \left( a + b + \sqrt{n} \left( \mn - \frac{c^2}{n} \right) \right)
% \NLeq
% %
% \half \left( a + b + \sqrt{n} \, \frac{n - c^2 + c^2}{n} \right) 
% \NLeq
% %
% \half \left( a + b + \sqrt{n} \, \frac{n}{n} \right) 
% \NLeq
% %
% \half \left( a + b + \sqrt{n} \right) 
% \NLeq
% %
\half \left( a + b + \sqrt{(b-a)^2 + c^2} \right).
\ea
Therefore we have that
\ba
\e{T_M - T_m} & \leq & \half \bigg( \e{T'} - \e{T} + \e{T} - \e{T'} 
\NL
+ \, \sqrt{ \left( \e{T} - \e{T'} + \e{T} - \e{T'} \right )^2 + 4 \left( \sqrt{\var{T}} + \sqrt{\var{T'}} \right) ^2 } \bigg)  
\NLeq
\sqrt{ \left( \e{T} - \e{T'} \right )^2 + \left( \sqrt{\var{T}} + \sqrt{\var{T'}} \right) ^2 }. 
\ea

For $T \neq T' $, we therefore propose
\ba
\lefteqn{\cov{H(T)}{H(T')}}
\NLeq
\sigma_1^2 \exp( \log \psi \, ( \e{T} + \e{T'} - 2 ) ) 
\NL
+ \, \sigma_R^2 \exp( \log \psi \, \sqrt{ \left( \e{T} - \e{T'} \right )^2 + \left( \sqrt{\var{T}} + \sqrt{\var{T'}} \right) ^2 } ).
\ea

% Issues as for the $J(T)$ case above are as follows:
% \bn
% \item When $\var{T} = \var{T'} = 0$, does the expression collapse to the known $\cov{H(t)}{H(t')}$ case?  No.
% \item The case when it is known that $T<T'$ doesn't seem very applicable here.
% \item What is $\var{H(T)}$?  For $J(T)$ it was very clear, and an indicator function could be used to expand the $\cov{J(T)}{J(T'}$ term to incorporate this.  Here, it isn't even clear what $\var{H(T)}$ should be.
% \en

% In my view, point (c) definitely needs addressing.  We need an expression for the variance, and it needs to coincide with the covariance expression when $T=T'$.  Point (a) may not need addressing - we could just accept that the conservative approximations/expression used above just don't resolve to the known input case, although clearly satisfy the inequalities.  Alternatively, we could make use of additional indicator functions when $\var{T} = \var{T'} = 0$ for the general covariance case and the case when $T=T'$ (i.e. recovering $\var{H(t)}$).

% So, $\var{H(T)}$ is not straightforward to access - it 
When $ T = T' $, the above expression needs to be commensurate with a reasonable expression for $ \var{H(T)} $.  Unlike $ \var{J(T)}$, we find that $ \var{H(T)} $ itself needs approximating. 
However, for a variance term we need to find an upper bound 
% (rather than a lower bound as for covariance) 
in order to be conservative with regard to prior variance specification (i.e. overestimate rather than underestimate).

% Finally, we might set
% \ba
% \lefteqn{\cov{H(T)}{H(T')}}
% \NLeq
% %
% \sigma_1^2 \exp( \log \psi \, ( \e{T} + \e{T'} - 2 ) ) 
% \NL
% %
% + \, \sigma_R^2 \exp( \log \psi \, \sqrt{ \left( \e{T} - \e{T'} \right )^2 + \left( \sqrt{\var{T}} + \sqrt{\var{T'}} \right) ^2 } )
% \ea

% Issues as for the $J(T)$ case above are as follows:
% \bn
% \item When $\var{T} = \var{T'} = 0$, does the expression collapse to the known $\cov{H(t)}{H(t')}$ case?  No.
% \item The case when it is known that $T<T'$ doesn't seem very applicable here.
% \item What is $\var{H(T)}$?  For $J(T)$ it was very clear, and an indicator function could be used to expand the $\cov{J(T)}{J(T'}$ term to incorporate this.  Here, it isn't even clear what $\var{H(T)}$ should be.
% \en

% In my view, point (c) definitely needs addressing.  We need an expression for the variance, and it needs to coincide with the covariance expression when $T=T'$.  Point (a) may not need addressing - we could just accept that the conservative approximations/expression used above just don't resolve to the known input case, although clearly satisfy the inequalities.  Alternatively, we could make use of additional indicator functions when $\var{T} = \var{T'} = 0$ for the general covariance case and the case when $T=T'$ (i.e. recovering $\var{H(t)}$).

% So, $\var{H(T)}$ is not straightforward to access - it needs approximating itself, however, for a variance term we need to find an upper bound (rather than a lower bound as for covariance) in order to be conservative (i.e. overestimate rather than underestimate).

We have that
\ba
\var{H(T)} & = & \e{\var{H(T)|T=t}} + \var{\e{H(T)|T}}
\NLeq
\e{\psi^{2(T-1)} \, \sigma^2_1 + \frac{1 - \psi^{2(T-2)}}{1 - \psi^2} \, \sigma_R^2}
\NLeq
\e{\psi^{2(T-1)} \, (\sigma_1^2 - \sigma_R^2) + \frac{1 - \psi^{2(T-1)}}{1 - \psi^2} \, \sigma_R^2}
\nonumber \\ & \leq &
\indicator_{\sigma_1^2 \leq K} K + (1 - \indicator_{\sigma_1^2 \leq K}) (\psi^4 \sigma_1^2 + \sigma_R^2),
\ea
where $K = \frac{\sigma_R^2}{1 - \psi^2}$.
The inequality holds because as $t \rightarrow \infty$, the expression tends to $K$, but this is from above or below depending on whether the variance gets larger or smaller for larger values of $t$.  We can assess this by considering the difference between the variance at time $t$ and $t+1$.
\ba
\var{H(t+1)} - \var{H(t)} & = & \psi^{2(t-1)} \sigma_R^2 - (\psi^{2(t-1)} - \psi^{2t}) \sigma_1^2
\NLeq
 \psi^{2(t-1)} \sigma_R^2 + (\psi^{2t} - \psi^{2(t-1)}) \sigma_1^2
 \NLeq
 \psi^{2(t-1)} (\sigma_R^2 + \psi^2 \sigma_1^2 - \sigma_1^2)
\ea
thus if $\sigma_R^2 + \psi^2 \sigma_1^2 - \sigma_1^2 \geq 0$, or in other words $\sigma_1^2 \leq K$, then the difference is positive, thus we approach the limit at infinity from below, and the expression is maximised at this limit.  On the other hand, if $\sigma_R^2 + \psi^2 \sigma_1^2 - \sigma_1^2 \leq 0$, or in other words $\sigma_1^2 \geq K$, then the difference is negative, and we approach the limit at infinity from above, in which case the expression is maximised for $t=3$ (since we enforced $t \geq 3$), which yields $\psi^4 \sigma_1^2 + \sigma_R^2$.

We can now set
\ba
\lefteqn{\cov{H(T)}{H(T')}}
\NLeq
\indicator_{T=T'} L + (1 - \indicator_{T=T'}) \bigg(  \sigma_1^2 \exp( \log \psi \, ( \e{T} + \e{T'} - 2 ) ) 
\NL
+ \, \sigma_R^2 \exp \left( \log \psi \, \sqrt{ \left( \e{T} - \e{T'} \right )^2 + \left( \sqrt{\var{T}} + \sqrt{\var{T'}} \right) ^2 } \right) \bigg)
\ea
where 
$$ L = \indicator_{\sigma_1^2 \leq K} K + (1 - \indicator_{\sigma_1^2 \leq K}) (\psi^4 \sigma_1^2 + \sigma_R^2)$$

% If we wish also to use indicators to enforce the case when $\var{T} = \var{T'} = 0$ to match the case of known $t,t'$, then we can have:
% \ba
% \lefteqn{\cov{H(T)}{H(T')}}
% \NLeq
% %
% \indicator_{T=T', \var{T} \neq 0} L + (1 - \indicator_{T=T', \var{T} \neq 0}) 
% \NL
% %
% \bigg(  \sigma_1^2 \exp( \log \psi \, ( \e{T} + \e{T'} - 2 ) ) + \, \sigma_R^2 \left( \frac{1 - \psi^{2(\e{T_m}-2}}{1 - \psi^2} \right)^{\indicator_{\var{T}=\var{T'}=0}}
% \NL
% %
% \qquad \times \, \exp \left( \log \psi \, \sqrt{ \left( \e{T} - \e{T'} \right )^2 + \left( \sqrt{\var{T}} + \sqrt{\var{T'}} \right) ^2 } \right) \bigg)
% \ea

% We can show this is appropriate since 
% \bi
% \item if $T=T'$ and $\var{T} \neq 0$, then we just get $L$, the term for $\var{T}$, T being random.
% \item if $\var{T} = \var{T'} = 0$, then the expression collapses as follows:
% \ba
% \lefteqn{\cov{H(T)}{H(T')}}
% \NLeq
% %
% \sigma^2_1 \exp( \log \psi (t + t' - 2) ) + \sigma_R^2 \left( \frac{1 - \psi^{2(t_m-2)}}{1 - \psi^2} \right) \exp( \log \psi \sqrt{(t-t')^2 + 0} )
% \NLeq
% %
% \sigma^2_1 \psi^{t + t' - 2} + \sigma_R^2 \left( \frac{1 - \psi^{2(t_m-2)}}{1 - \psi^2} \right) \psi^{|t-t'|} 
% \NLeq
% %
% \psi^d \left( \sigma^2_1 \psi^{2(t_m - 1)} + \sigma_R^2 \left( \frac{1 - \psi^{2(t_m-2)}}{1 - \psi^2} \right) \right)
% \ea
% which itself naturally collapses appropriately when $t=t'$ to
% \be
% \var{H(t)} =  \sigma^2_1 \psi^{2(t - 1)} + \sigma_R^2 \left( \frac{1 - \psi^{2(t-2)}}{1 - \psi^2} \right) 
% \ee
% \item The remaining cases are as given with both indicator function values set to $0$.
% \ei

Finally, to conclude our prior specification in the case of uncertain inputs for the model presented in this section, note that, with $\bT = (1,T)^T$, we have
\ba
\lefteqn{\e{\bT^T\bDe \bT'}}
\NLeq
\trace( \bDe (\e{\bT'} \e{\bT^T} + \cov{\bT}{\bT'} ) )
\NLeq
\trace \left( 
\left( \begin{array}{cc} \delta_{00} & \delta_{01} \\ \delta_{10} & \delta_{11} \end{array} \right) 
\left( \begin{array}{cc} 1 & \e{T} \\ \e{T'} & \e{T} \e{T'} \end{array} \right) 
+
\left( \begin{array}{cc} 0 & 0 \\ 0 & \cov{T}{T'} \end{array} \right) 
\right)
\NLeq
\trace \left( \begin{array}{cc} \delta_{00} + \delta_{01} \e{T'} & \delta_{00} \e{T} + \delta_{01} \e{T} \e{T'} \\ \delta_{10} + \delta_{11} \e{T'} & \delta_{10} \e{T} + \delta_{11} \e{T} \e{T'} + \cov{T}{T'} \end{array}   \right)
\NLeq
\delta_{00} + \delta_{01} ( \e{T'} + \e{T} ) + \delta_{11} \e{T} \e{T'} + \cov{T}{T'},
\ea
and
\be
\bGa^T \cov{\bT}{\bT'} \bGa = 
\left( \begin{array}{cc} \gamma_0 & \gamma_1 \end{array} \right)
\left( \begin{array}{cc} 0 & 0 \\ 0 & \cov{T}{T'} \end{array} \right)
\left( \begin{array}{c} \gamma_0 \\ \gamma_1 \end{array} \right)
= \gamma_1^2 \cov{T}{T'},
\ee
so that full prior specification for $y(T)$ can therefore be given by
\be
\e{y(T)} = \gamma_0 + \e{T} \gamma_1,
\ee
and 
\ba
\lefteqn{\cov{y(T)}{y(T')}}
\NLeq
\delta_{00} + \delta_{01} ( \e{T'} + \e{T} ) + \delta_{11} \e{T} \e{T'} + \cov{T}{T'} 
\NL
+ \,\, \gamma_1^2 \cov{T}{T'} 
\NL
+ \,\, \indicator_{T=T'} \sigma_A^2
\NL
+ \,\, \frac{\sigma^2_Q}{2} \left( \e{T} + \e{T'} - \sqrt{ \left( \e{T'} - \e{T} \right)^2 + \left( 1 - \indicator_{T=T'} \right) \left (\sqrt{\var{T}} + \sqrt{\var{T'}} \right)^2} \right)
\NL
+ \,\, \indicator_{T=T'} L + (1 - \indicator_{T=T'}) \bigg(  \sigma_1^2 \exp( \log \psi \, ( \e{T} + \e{T'} - 2 ) ) 
\NL
\qquad + \sigma_R^2 \exp \left( \log \psi \, \sqrt{ \left( \e{T} - \e{T'} \right )^2 + \left( \sqrt{\var{T}} + \sqrt{\var{T'}} \right) ^2 } \right) \bigg).
\ea

%%%%%%%%%%%%%%%%%%%%%%%%%%%%%%%%%%%%%%%%%%%%%%

\section{Proof of \sam{Lemmas 1 - 4} \label{LemmaProofs}}

In this section, we prove Lemmas 1 - 4
of Sections 4 and 5 of the main text.

%%%%%%%%%%%%%%%%%%%%%%%%%%%%%%%%%

\subsection*{Proof of Lemma 1}

Rewrite Equation (60) as
\be \nonumber
\begin{split}
c(\bX,\bX') & = \exp \left\{ - \e{(\bX - \bX')^T\Theta^{-2}(\bX - \bX')}\right\}\\
&= \exp \left\{ - \e{\bX^T\Theta^{-2}\bX} \right\} \exp \left\{ -\e{(\bX')^T\Theta^{-2}\bX'} \right\} \exp \left\{ 2\e{\bX^T\Theta^{-2}\bX'} \right\}\,, \\
\end{split}
\ee
with kernel $$ k(\bX , \bX') = \exp \left\{ 2\e{\bX^T\Theta^{-2}\bX'} \right\} \, . $$
The kernel $ k(\bX, \bX') $ is positive definite since 
$$ \mathrm{tr}(\Theta^{-2}\var{\bX}) + \e{\bX}^T\Theta^{-2}\e{\bX} \ge 0 $$ 
with equality if and only if $\bX$ has a degenerate distribution at zero. Postive definiteness of the function $c(\bX, \bX')$ then follows from standard properties of kernels (see \citealp{stc2011}, ch.~3)

%%%%%%%%%%%%%%%%%%%%%%%%%%%%%%%%%

\subsection*{Proof of Lemma 2}

Covariance can be derived from conditional quantities using the law of total covariance, hence:
\begin{multline*}
\cov{\uX}{\uXp} = \e{\cov{\uX}{\uXp\,|\,\bX = \bx, \bX'=\bx'}} \\ 
+ \cov{\e{\uX\,|\, \bX}}{\e{\uXp\,|\, \bX'}}\,.
\end{multline*}
Under the assumption that $\e{\bu(\bX)} = 0$, it follows that 
$$\cov{\e{\uX\,|\, \bX}}{\e{\uXp\,|\, \bX'}} = 0$$. 
Hence for conditional covariance (61), we have
\begin{equation*}
\begin{split}
\cov{\uX}{\uXp} & = \mathrm{E}\left[
\exp \left( - \sum_{r=1}^p \left\{ \frac{X_{(r)} - X'_{(r)}}{\theta_r} \right\}^2  \right)
\right] \, \Sig \\
& \succeq  \exp \left( 
- \sum_{r=1}^p \mathrm{E}\left[
\left\{ \frac{X_{(r)} - X'_{(r)}}{\theta_r} \right\}^2 \,
\right]
\right) \, \Sig \, , \\
\end{split}
\end{equation*}
under the Loewner (partial) ordering, following from an application of Jensen's inequality and the positive-definiteness of $\Sig$. As the diagonal entries of $\Sig$ are all non-negative, elementwise inequality for the $kk$th entry follows directly ($k = 1, \ldots, q$).

%%%%%%%%%%%%%%%%%%%%%%%%%%%%%%%%%

\subsection*{Proof of Lemma 3}

\sam{We begin by expanding the terms of the Bayes linear update as follows:}
\ba
\eF{\fX} & = &  \eF{\wX \, \bb}   +   \eF{\uX}\,.  \nonumber
\ea
Taking the two parts of the right-hand side of this equation separately, we first have that
\be
\eF{\wX \, \bb} = \eF{\wX} \, \eF{\bb} = \e{\wX} \, \eF{\bb}\,,  \label{eFwXb} 
\ee
where we have used the facts that $ \eF{\wX} = \e{\wX} $ since $ \cov{\bX}{\bF} = \bzero $, and $ \covF{\wX}{\bb} = \bzero $ since $ \covF{\gX}{\bb} = \bzero $.
\sam{We then have, using basic rules of linear algebra \citep{Mardia79_MVA}, that
\ba
\eF{\uX} & = & \e{\uX} + \cov{\uX}{\bF} \, \var{\bF}^{-1} \, ( \bF - \e{\bF} )
\NLeq
\cov{\uX}{\bU} \, \var{\bF}^{-1} \, (\bF - \bW \, \bGa )
\NLeq
\vX \, ( \WDW + \bO )^{-1} \, ( \bF - \bW \, \bGa )
\NLeq
\vX \, \big( ( \WDW + \bO )^{-1} \, \bF - ( \WDW + \bO )^{-1} \, \bW \, \bGa \big)
\NLeq
\vX \bigg( \big( \bOi - \bOi \, \bW \, ( \bDi + \WOW )^{-1} \, \bW^T \, \bOi \big) \, \bF 
\NL
- \bOi \, \bW \, ( \WOW + \bDi )^{-1} \, \bDi \, \bW \bigg)  
\NLeq
\vX \, \bOi \, \big( \bF - \bW \, ( \bDi + \WOW )^{-1} \, ( \bDi \, \bW + \bW \, \bOi \, \bF ) \big) 
\NLeq
\vX \, \bOi \, ( \bF - \bW \, \eF{\bb} )\,,   \label{eFuX}
\ea
where we have that 
\ba
\eb & = &  \e{\bb} + \cov{\bb}{\bF} \var{\bF}^{-1} (\bF - \e{\bF} )   
\NLeq
\bGa   +   \bDe \, \bW^T \, ( \WDW + \bO )^{-1} ( \bF - \bW \bGa )
\NLeq
\bGa   +   ( \WOW + \bDi )^{-1} \WO ( \bF - \bW \bGa )
\NLeq
( \WOW + \bDi )^{-1} \big( ( \WOW + \bDi ) \bGa    +    \WO ( \bF - \bW \bGa ) \big)
\NLeq
( \WOW + \bDi )^{-1} ( \DG + \WO \, \bF )
\NLeq
( \WOW + \bDi )^{-1} ( \WOW \, \bbgls   +   \DG )
\ea
where line 3 holds as a result of the identity (see, for example, \cite{Mardia79_MVA})
\begin{equation}
AB(DAB + C)^{-1}  =  (BC^{-1}D + A^{-1})^{-1}BC^{-1} \label{MI1}
\end{equation}
where here, $A, B, C, D$ are generic matrices, and
\ba
\bbgls & = & ( \WOW )^{-1} \WO \, \bF
\NLeq
\identity_q \otimes ( \GCG )^{-1} \GC \, \bF  \label{bbgls}
\ea
is the generalised least squares estimate for $ \bb $. 
Combining Equations \eqref{eFwXb} and \eqref{eFuX} we get:
}
\ba
\eF{\fX} & = & \ewX \, \eb   +    \vX \, \bOi \, (\bF - \bW \, \eb )\,. \nonumber
\ea
\begin{flushright} $ \Box $ \end{flushright}

%%%%%%%%%%%%%%%%%%%%%%%%%%%%%%%%%

\subsection*{Proof of Lemma 4}

\sam{We begin by expanding the terms of the Bayes linear update as follows:}
\ba
\varF{\fX} & = & \varF{\wX \, \bb + \uX}   
\NLeq
\varF{\wX \, \bb} + \varF{\uX}  
\NL 
\, + \, \covF{\wX \, \bb}{\uX} + \covF{\uX}{\wX \, \bb}\,.    \label{var_gZ} 
\ea
%%%
\sam{Starting with the first term on the right-hand side of Equation \eqref{var_gZ}, we have, using linear algebra \citep{Mardia79_MVA}:}
\ba
\varF{ \wX \, \bb } & = & \e{ \varFX{ \wX \, \bb } } + \var{ \eFX{ \wX \, \bb } }
\NLeq
\e{ \wX \, \vb \, \wX^T }    +    \ebt \, \var{ \wX }  \eb \,,   \label{varFwXb}
\ea
where the first term of Equation \ref{varFwXb} can be calculated component-wise:
\ba
\ed{ij}{\wX \vb \wX^T} & = & \e{ \trace ( \wix \, \vb \, \wjx^T ) }
\NLeq
\e{ \trace( \vb \, \wjx^T \, \wix ) }
\NLeq
\trace( \e{ \vb \, \wjx^T \, \wix } )
\NLeq
\trace( \vb \e{\wjx^T \, \wix} )
\\ & = &
\trace \big( \vb ( \e{ \wjx^T} \e{ \wix } ) + \cov{\wjx}{\wix}    \big) \nonumber
\ea
which is now in terms of specified quantities.  Note that $\vb$ is given by
\ba
\vb & = & \var{\bb} - \cov{\bb}{\bF} \var{\bF}^{-1} \cov{\bF}{\bb}
\NLeq
\bDe   -    \bDe \, \bW^T (\WDW + \bO)^{-1} \bW \, \bDe
\NLeq
\bDe    -   (\WOW + \bDi)^{-1} \WOW \, \bDe
\NLeq
(\WOW + \bDi)^{-1} \big( (\WOW + \bDi) \bDe - \WOW \, \bDe \big)
\NLeq
(\WOW + \bDi)^{-1}
\ea
For the second term of Equation \eqref{var_gZ}, we have that
%%%
\sam{
\ba
\lefteqn{ \varF{ \uX } } \nonumber \\ 
& = &  \var{ \uX } -  \cov{\uX}{\bF} \, \var{\bF}^{-1} \, \cov{\bF}{\uX}
\NLeq
\Sig - \vX \, ( \WDW + \bO )^{-1} \, \vX^T
\NLeq
\Sig - \vX \, ( \bOi - \bOi \, \bW \, ( \bDi + \WOW )^{-1} \, \bW^T \, \bOi ) \, \vX^T
\NLeq
\Sig - \vX \, \bOi \, \vX^T   +   \vX \, \bOi \, \bW \, \vb  \, \bW^T \, \bOi \, \vX^T \,,  \label{varFuX}
\ea
}
and for the third term of Equation \eqref{var_gZ}, we have that
%%%
\sam{
\ba
\lefteqn{\covF{ \wX \, \bb }{ \uX }} \nonumber \\ 
& = & \cov{ \wX \, \bb }{\uX} - \cov{ \wX \, \bb }{\bF} \var{\bF}^{-1} \cov{\bF}{\uX} \, . \label{thirteen}
\ea
In Equation \eqref{thirteen}, we have that:
\ba
\cov{ \wX \, \bb }{\bF} & = & \cov{ \wX \, \bb}{ \bW \,\bb}
\NLeq
\cov{ \wX \, \bb}{ \bb} \bW^T
\NLeq
( \e{ \wX \, \bb \, \bb^T } - \e{ \wX \, \bb } \e{ \bb^T } ) \bW^T
\NLeq
( \e{ \wX } \e{ \bb \, \bb^T } - \e{ \wX } \e{ \bb } \e{ \bb^T } ) \bW^T
\NLeq
\e{ \wX } \var{\bb} \bW^T
\NLeq
\e{ \wX } \bDe \bW^T\,, \nonumber
\ea
}
so that
\ba
\covF{ \wX \, \bb}{\uX} & = & - \, \e{\wX} \, \bDe \, \bW^T (\WDW + \bO)^{-1} \vX^T
\NLeq
- \, \e{\wX} ( \WOW + \bDi )^{-1} \bW \, \bOi \, \vX^T
\NLeq
- \, \e{\wX} \vb \WO \vX^T \,.  \label{fiftenn}
\ea
\sam{Putting Equations \eqref{varFwXb}, \eqref{varFuX} and \eqref{fiftenn} together, we get that:}
\ba
\varF{ \fX } & = & \e{ \wX \, \vb \, \wX^T }   +   \ebt \, \vwX \, \eb + \, \Sig  
\NL
\,    - \,   \vX \, \bOi \, \vX^T    +    \vX \, \bOi \, \bW \, \vb  \, \bW^T \, \bOi \, \vX^T
\NL
\, - \, \ewX \, \vb \, \bW \, \bOi \, \vX^T    
\NL
\, - \,   ( \ewX \, \vb \, \bW \, \bOi \, \vX^T ) ^T.  \label{UIBLE_lem_Var}
\ea
\begin{flushright} $ \Box $ \end{flushright}

\end{document}